\documentclass[fleqn,usenatbib]{mnras}
\usepackage{newtxtext,newtxmath}

\usepackage[T1]{fontenc}
\usepackage{ae,aecompl}

\usepackage{graphicx}	
\usepackage{amsmath}	
\usepackage{xspace}
\usepackage{amsfonts,textcomp}
\usepackage[usenames]{color}
\usepackage{times}
\usepackage{hyperref}

\newcommand{\be}{\begin{equation}}
\newcommand{\ee}{\end{equation}}

\newcommand{\msun}{\rm M_{\sun}}

\newcommand{\nh}{n_{\rm H}}

\newcommand{\K}{\rm K}

\newcommand{\HI}{\ion{H}{i}}
\newcommand{\HH}{\mbox{H$\rm _2$}}

\newcommand{\CO}{\mbox{ CO }}

\definecolor{darkgreen}{rgb}{0.0,0.5,0.0}
\definecolor{darkred}{rgb}{0.5,0.0,0.0}
\definecolor{brown}{rgb}{0.65,.16,0.246}
\definecolor{grey}{rgb}{0.4,0.5,0.6}

\title[HDGAS; Molecular Gas Dynamics]{Hydrodynamic simulations of the Disk of Gas Around Supermassive black holes (HDGAS) -I; Molecular Gas Dynamics}
\author[M. Raouf et al.]{
Mojtaba~Raouf,$^{1}$\thanks{E-mail: *mojtaba.raouf@gmail.com, raouf@strw.leidenuniv.nl} 
 Serena Viti,$^{1,6}$ 
 S. Garc{\'\i}a-Burillo,$^{2}$
 Alexander J. Richings,$^{3,4}$
  \newauthor
 Joop schaye,$^{1}$
 Ashley~Bemis,$^{1}$
 Folkert~S.J.~Nobels,$^{1}$
 Matteo~Guainazzi,$^{5}$
 Ko-Yun~Huang,$^{1}$
 \newauthor
 Matthieu~Schaller,$^{1}$
 Violette Impellizzeri,$^{1}$
 Jon Holdship,$^{1,6}$\\
$^{1}$ Leiden Observatory, Leiden University, P.O. Box 9513, 2300 RA Leiden, Netherlands\\
$^{2}$ Observatorio Astronómico Nacional (OAN-IGN)-Observatorio de Madrid, Alfonso XII, 3, 28014 Madrid, Spain\\
$^{3}$ E. A. Milne Centre for Astrophysics, Department of Physics and Mathematics, University of Hull, Cottingham Road, Hull, HU6 7RX, UK\\
$^{4}$ DAIM, University of Hull, Cottingham Road, Hull, HU6 7RX, UK\\
$^{5}$ ESA/ESTEC, D-SRE, Keplerlaan 1, NL-2200 AG, Noordwijk, Netherlands\\
$^{6}$ Department of Physics and Astronomy, University College London, Gower Street, WC1E 6BT, London, UK\\
}
\begin{document}
\label{firstpage}
\pagerange{\pageref{firstpage}--\pageref{lastpage}}
\maketitle
\begin{abstract}
We present hydrodynamic simulations of the interstellar medium~(ISM) within the circumnuclear disk~(CND) of a typical AGN-dominated galaxy influenced by mechanical feedback from an active galactic nucleus~(AGN). The simulations are coupled with the CHIMES non-equilibrium chemistry network to treat the radiative-cooling and AGN-heating. A focus is placed on the central 100~pc scale where AGN outflows are coupled to the ISM and constrained by observational Seyfert-2 galaxies. AGN-feedback models are implemented with different wind-velocity and mass-loading factors. We post-process the simulation snapshots with a radiative-transfer code to obtain the molecular emission lines. We find that the inclusion of an AGN promotes the formation of CO in clumpy and dense regions surrounding supermassive-blackholes~(SMBH). The CO(1-0) intensity maps~($<$6~Myr) in the CND seem to match well with observations of NGC~1068 with a best match for a model with 5000~$\rm km/s$ wind-velocity and a high mass-loading factor.  We attempt to discern between competing explanations for the apparent counter-rotating gas disk in the NGC~1068 through an analysis of kinematic maps of the CO line emission.  We suggest that mechanical AGN-feedback could explain the alignment-stability of position-angle across the different CND radii around the SMBH through momentum and energy loading of the wind. It is the wind-velocity that drives the disk out of alignment on a 100 pc scale for a long period of time. The position-velocity diagrams are in broad agreement with the predicted Keplerian rotation-curve in the model without-AGN, but the AGN models exhibit a larger degree of scatter, in better agreement with  NGC~1068 observations.

\end{abstract}


\section{Introduction}

Supermassive black holes (SMBHs) at the center of  massive galaxies provide a mechanism for explaining the existence of Active Galactic Nuclei (AGN),
which inhibit the formation of excessive stars and, therefore, the growth of galaxies \citep{Silk1998}. Radiative and mechanical feedback from AGN has been suggested as a mechanism for galaxy evolution and self-regulation in a variety of theoretical models and numerical simulations \citep[e.g][]{DiMatteo2005,DiMatteo2008,Croton2006,Booth2009,Dubois2013,Bower2017,Raouf2017,Raouf2019,Dave2019}. Hence, understanding the origin and sustainment of AGNs in galaxies requires an investigation of the dynamics of the interstellar gas surrounding SMBHs.

Since most of the gas in the centre of galaxies is in molecular form \citep[often extended well beyond  1 Kpc, e.g.][]{FriasCastillo2022}, submillimeter molecular tracers have now become  the most important probes of AGNs. However, 
in the last decade molecular observations of the gas surrounding AGNs has revealed a complex picture whereby mechanical and radiative feedback observational signatures are complicated by the likely presence of starbursts \citep[e.g.][]{Aalto2012,Martin2011,Aladro2013,Watanabe2014}. Nevertheless, it has been shown that - for the nearest sources - multi-line multispecies observations allow us to study in detail the different gas components surrounding the AGN \citep[e.g.][]{Meier2005,Garcia-Burillo2010}.
Galaxies characterized by both an AGN and strong circumnuclear star formation are of particular interest, because they present a complex combination of energetic processes (i.e gas accretion, external infall  etc...), as well as (and also giving rise to) a complex geometry and kinematics \citep{Raouf2018,Raouf2021,Knapen2019,Winkel2022}. 
 \begin{figure}
	\centering
        \includegraphics[width=0.9\linewidth]{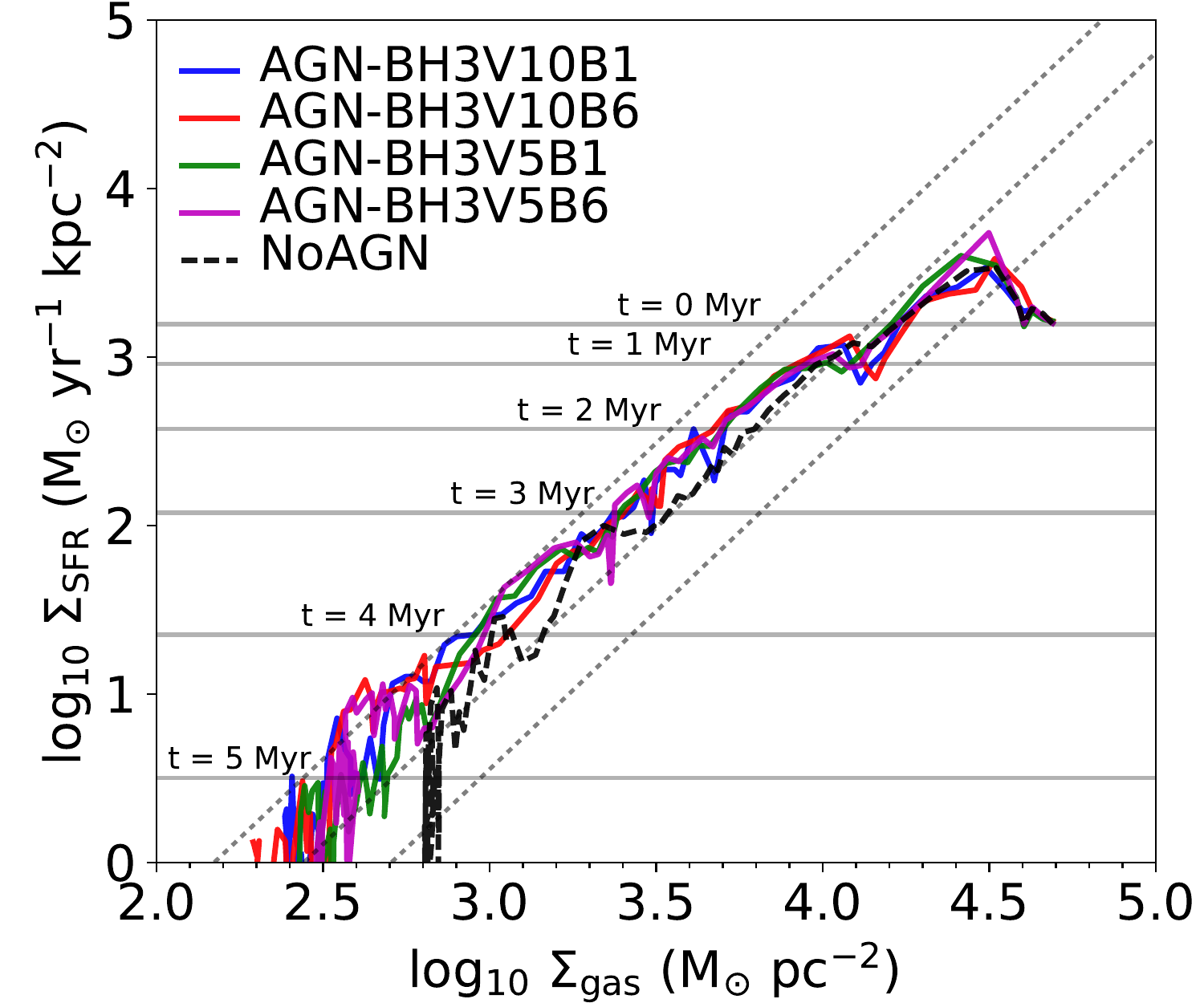}
	\caption{
  A comparison of the simulations disk (see Table \ref{tab:AGN_table}) conducted at all times (10 Myrs; Grey horizontal lines represent some specific times) on the Kennicutt-Schmidt relations. We have calculated the surface densities of gas and star formation rates within 100 pc and determined whether they are consistent with the observational a range. This observations are from range of galaxies (not limited to galactic nuclei) with a scatter of $\pm 0.5$~dex as shown by the grey dotted lines \citep{Narayanan2012}.
 }
\label{fig:KS}
\end{figure}

The best studied example is NGC~1068, a nearby Seyfert type 2 AGN galaxy, with a massive black hole of 8$\times$10$^6$ M$_{\odot}$\citep{Lodato2003} and a  circumnuclear disk (CND) of about 200 pc in size. The many high spatial resolution molecular observations \citep[e.g.][]{Garcia-Burillo2014,Imanishi2018, Garcia-Burillo2016,Viti2014,Scourfield2020,Imanishi2020} of NGC~1068 reveal a complex structure and dynamics with: (i) a massive thick gas disk seen edge-on, obscuring the central engine, 
with a radius of $\sim$ 10~pc \citep{Garcia-Burillo2016,Imanishi2016, Imanishi2018}; (ii) a counter-rotating inner disk (torus) at radii smaller than $\sim$ 1.2 pc \citep{Impellizzeri2019}; (iii) a molecular outflow in the inner region of the disk ($\leq$ 3~pc) \citep{Gallimore2016,Garcia-Burillo2019,Impellizzeri2019}. The torus is connected to the CND via a network of gas lanes whose kinematics are consistent with a 3D outflow geometry \citep{Garcia-Burillo2019}; and  (iv)  a starburst ring (SB ring, r $\sim$ 1.3~kpc).

The overall dynamics of the CND in NGC~1068 will be the result of the sum of several processes at play:  accretion of gas within the disk, external infall, AGN feedback, and interaction of the CND gas with the surrounding starburst. A possible footprint of AGN feedback may be found in the dramatic changes in the molecular  line ratios across the CND, which are tightly associated with UV and X-ray illumination of the region by the AGN. \citet{Garcia-Burillo2014} found that the CO gas was being driven by an AGN, with an outflow rate of $63_{-37}^{+21}\ \rm M_{\odot} yr^{-1}$ from the CND, giving a timescale for gas depletion of $\leq$ 1 Myr. In fact, \citet{Garcia-Burillo2019} showed that there is a deficiency of molecular gas in the central $\simeq100$ pc of the CND. This region is pervaded by strong emission from a wide-angle bicone of ionized gas \citep{Das2006, MuellerSanchez2011, Barbosa2014}. By analyzing the distortions of the molecular gas velocity field using ALMA with a spatial resolution of 35 pc \citet{Garcia-Burillo2014} concluded that the ionized outflow is responsible for initiating a molecular outflow across a region extending from 50 pc out to 500 pc.  However, efficient gas inflow from the SB ring may refresh the reservoir over a much longer period of time. Depending on the physical processes and factors at play, some observations suggest that the ionized gas outflow in NGC~1068 is slowing down and showing up in the molecular gas phase. For instance, \cite{Shin2021} showed that although NGC 1068 is known to have gas outflows originating from its AGN, more complex kinematical signatures have recently been detected that are in conflict with rotating outflows or simple biconical outflows.

 \begin{figure*}
	\centering
         \includegraphics[width=0.45\linewidth]{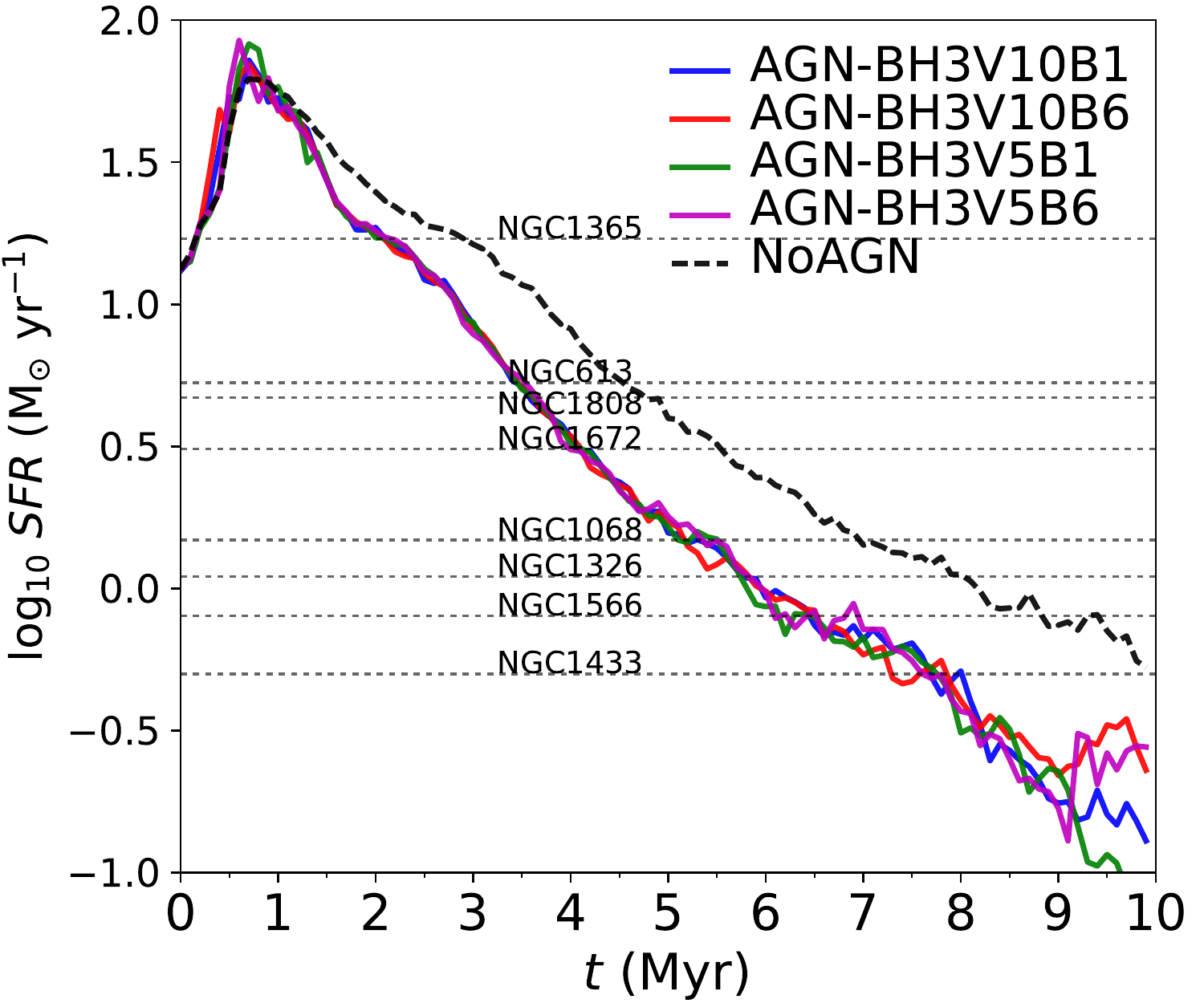}
         \includegraphics[width=0.45\linewidth]{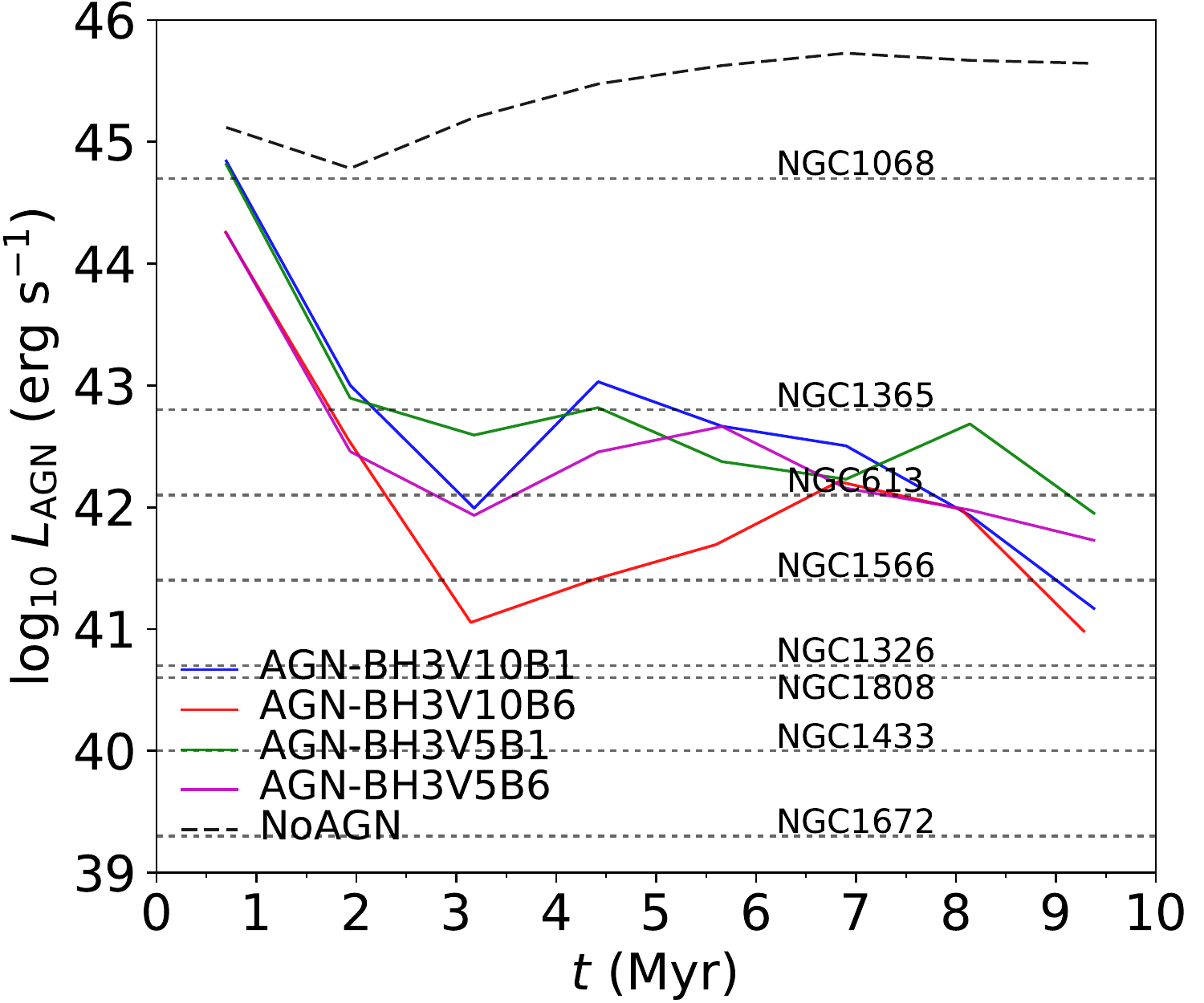}
	\caption{Left: Evolution of the total SFR within the central r=200~pc region for simulations with various AGN feedback models indicate by colors and the model NoAGN (black dashed line). Black dotted lines indicate the median SFR of Seyfert galaxies like NGC~1068 for a variety of spatial resolutions \citep[see Tab. 1][]{Combes2019}.
 Right: Evolution of AGN luminosity for different AGN models and model NoAGN. Observed AGN luminosities of Seyfert galaxies are show by dotted lines \citep[see Table 7][and ref. therein]{Combes2019}. The AGN luminosity of NGC~1068 is around $10^{44.7} \rm erg\ s^{-1}$\citep{Gallimore2001} 
 }
\label{fig:SFR}
\end{figure*}

\citet{Vollmer2022} applied the analytical model of a turbulent clumpy gas disk of \citet{Vollmer2013} to produce an NGC~1068 dynamical model that could reproduce the high spatial resolution ALMA CO, HCO$^+$ and HCN observations of the nucleus of NGC~1068. They find that the simplest configuration that can explain the kinematics of NGC~1068 is that of infalling gas clouds into a preexisting gas ring and that the best fit cloud-ring collision is prograde for the CND and retrograde for the NGC~1068. The collision of the cloud that falls inwards increases the radial inflow of gas, which converts potential energy into turbulence as it moves closer to the center. They also find that turbulent mechanical heating is the dominant mechanism within the thick gas disk. This might be reflected in the distinct values of momentum and energy carried by the AGN wind. However their model lacked of a proper 3D treatment of the radiative transfer for the molecular line emission. 

Taking a Seyfert-2 type galaxy (e.g NGC~1068) as our template, in this study we present a model of a gas disk around a black hole using  hydrodynamical simulations that incorporate direct radiative cooling, non-equilibrium chemistry, and mechanical feedback by the central supermassive black hole and we directly compare this model to one without AGN feedback. A post-processing radiative transfer code is also used to compare different CO emission lines moments (0, 1, 2) maps.  
The aim of this paper is to study in detail how the central supermassive black hole impacts the  molecular gas dynamics, and whether AGN feedback has any implications for the kinematic properties of a disk through the analysis of the position angles of disks surrounding supermassive black holes with different mechanical feedback energy and momentum loading factors over time.

The structure of the paper is as follows: the simulation and methods are described in Section \ref{sec:simulation}.  In Section \ref{sec:results1}, we present our results and analysis. Finally, we provide a summary and discussion in Section \ref{subsec:sum}. 

\begin{figure*}
	\centering
	\includegraphics[width=1.0\linewidth]{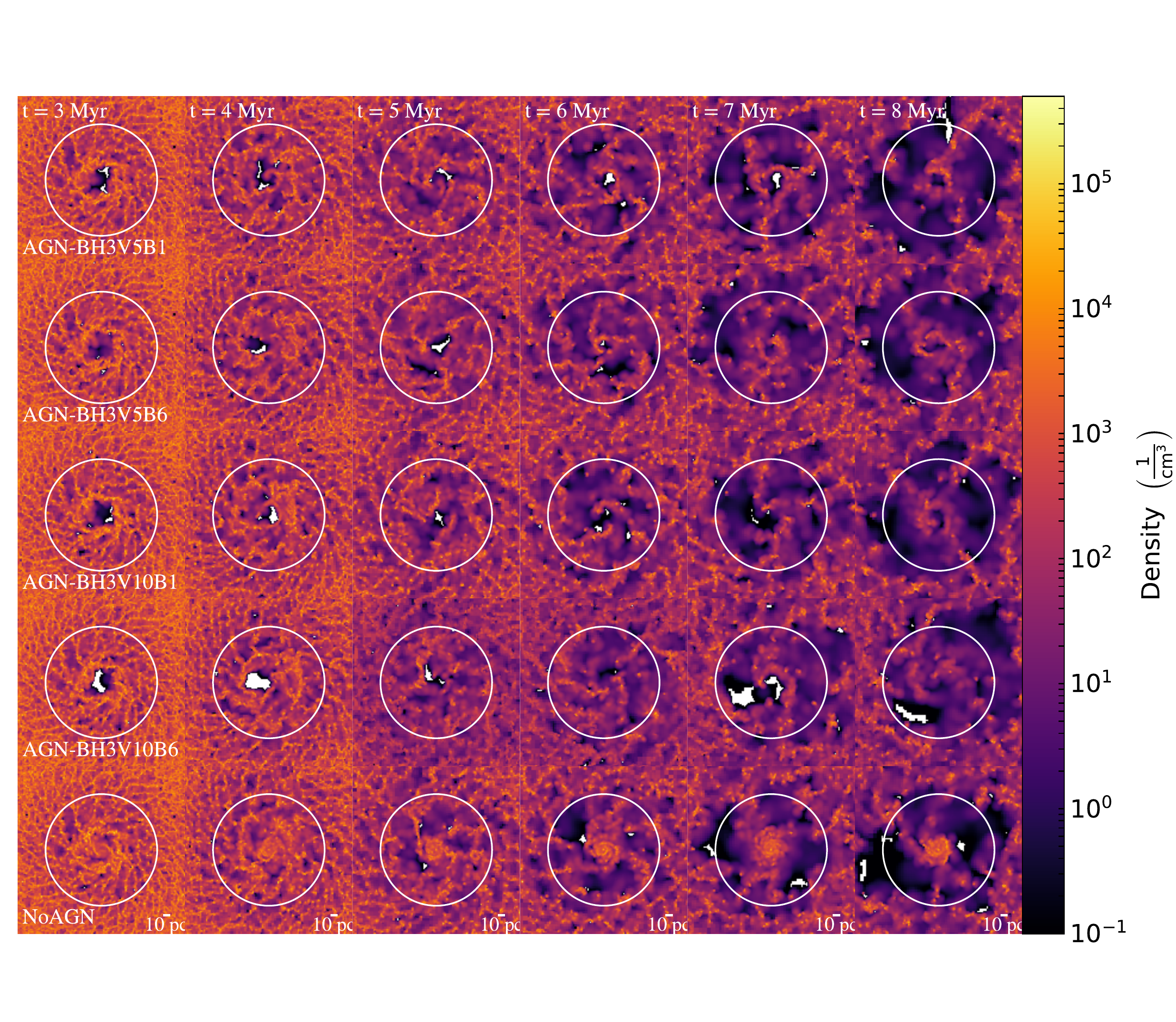}
	\caption{ Face-on projection of the volume density for the central 150 pc of the disk radius for different times between 3 and 8 Myr (left to right) of different AGN models and NoAGN feedback (top to bottom). The central r=100~pc scale is shown by a circle in each panel.
 }
\label{fig:Map_density}
\end{figure*}

\begin{figure*}
\centering
 \includegraphics[width=1.0\linewidth]{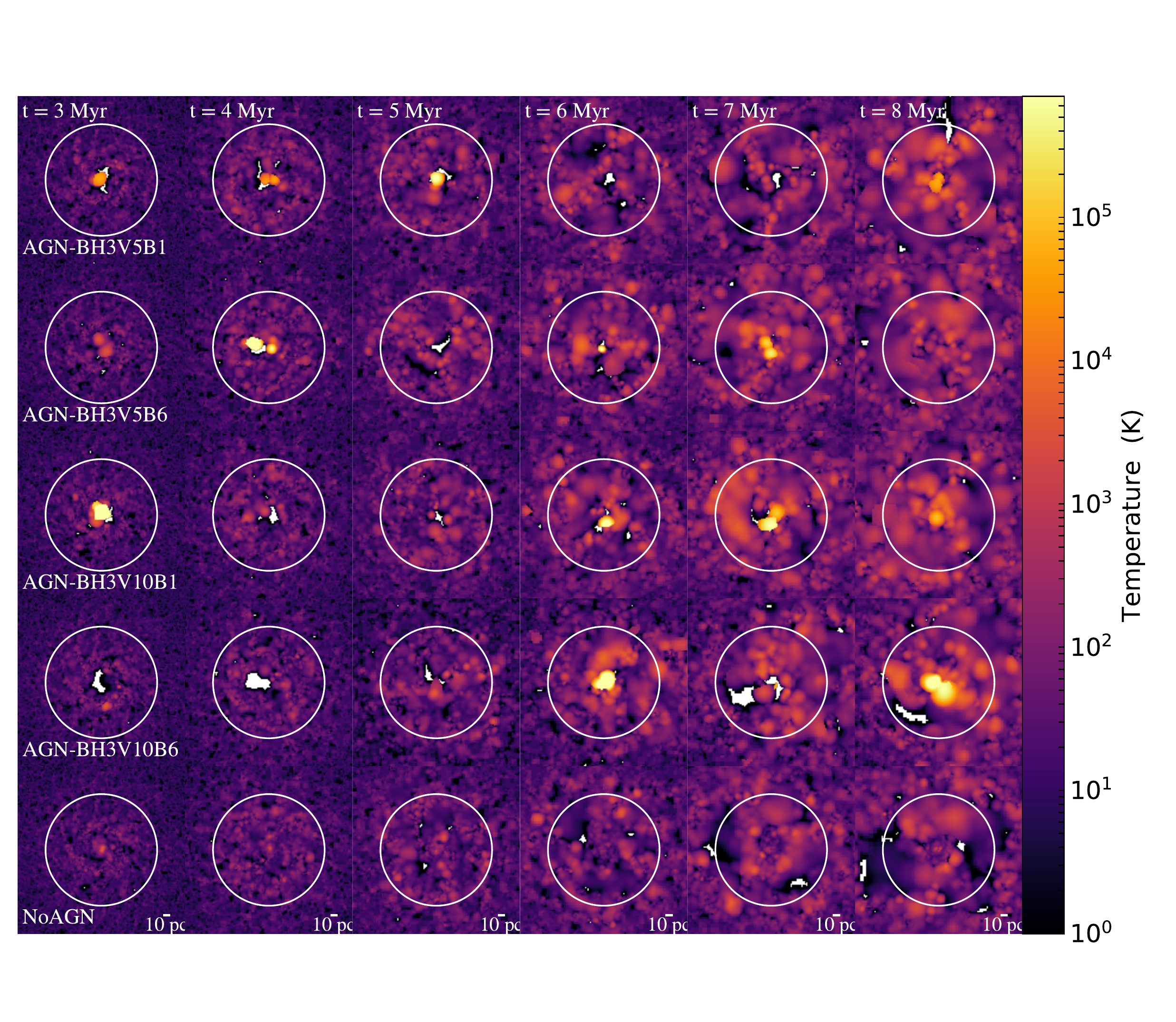}
 \caption{ Same as figure \ref{fig:Map_density} but for the gas temperature.
 }
\label{fig:Map_Temp}
\end{figure*}

\section{The Simulations} \label{sec:simulation}

 In this study, the simulations were run using Gizmo \citep{Hopkins2015}, a hydrodynamics and gravity code that implements several different hydrodynamics solvers. For our simulations, we use the Pressure-Entropy formulation of the Smoothed Particle Hydrodynamics method (P-SPH). A significant benefit of P-SPH is that it improves gradient calculations and artificial viscosity to capture interfaces and instabilities, as well as  the better treatment of contact discontinuities \citep{Hopkins2015}.

 An initial condition is a gas-rich nuclear disk containing a black hole in its center created using the code ``MakeNewdisk" \citep{Springel2005a}. 
 We incorporated the gravitational potential from an exponential nuclear disk of gas (scale-length $h=100\,$pc and mass of $M_{D}\simeq10^{8}\,\msun$) with vertical hydrostatic gas equilibrium and the central super massive black hole (SMBH), using a single collisionless particle with a mass of $M_{\rm{BH}} = 10^{7} \rm{M}_{\odot}$ \citep{Lodato2003}. 
 In this simulation, we do not include any stellar components or DM haloes. 
 Our SMBH mass is derived using specific observational measurements using as a template  the galaxy NGC~1068,  rather than the $M_{\rm{BH}} - \sigma$ \citep{Haring2004,McConnell2013} relation. In our simulations we do not constrain the BH particle to be fixed at the centre of the potential, as the mass of the black hole exceeds that of the other baryonic particles by many orders of magnitude, so it naturally remains near the centre. 
 
 The initial gas disk contains $10^6$ particles, where each particle initially has a mass of  $\sim100\ \msun$.
 Self-gravity of the gas is also included. We use adaptive gravitational softening lengths for gas particles with minimum smoothing length of 0.1 pc and a constant gravitational softening length of 1 pc for all star and BH particles.

We have calculated that the average free-fall time ($t_{\rm ff}$) for our disk is 0.25 Myr, which means that a simulation time of 10 Myr (equivalent to 40 $t_{\rm ff}$) is appropriate. Additionally, observations have shown that the gas depletion timescale is less than 1 Myr \citep{Garcia-Burillo2014}, making a simulation time of 10 Myr ideal. To ensure stability, we have conducted idealised tests using only hydrodynamics and gravity, which have confirmed that the setup is stable for the duration of the simulation. Our simulations model a disk with a scale length of 100 pc and are contained within a box measuring $(1~\rm kpc)^3$. We have set the initial temperature and metallicity of the disk to $10^4~\rm K$ and 100 per cent solar \citep[$Z_{\odot}=0.0196$][]{vonSteiger2016}, respectively.

\subsection{Radiative cooling and chemistry}\label{sec:cooling}

With the \textsc{chimes} non-equilibrium chemistry and cooling model \citep{Richings2014a,Richings2014b}, chemical abundances, for 157 species  are calculated over time,  including all ionization states of 11 elements that are important for cooling, as well as for 20 molecules, including  \HH~ and \CO. By integrating the temperature in time with the 157 rate equations, the \textsc{chimes} module calculates cooling and heating rates. Chemical reactions in the \textsc{chimes} chemical network include collisional ionization, recombination, charge transfer, and reactions on the surface of dust grains, such as the formation of molecular hydrogen (assuming a constant ratio of dust to metals) as well as  photoionisation and photodissociation. This collection of chemical reactions in \textsc{chimes} can be found in table~B1 of \citet{Richings2014a}. In this model, the cooling curves for the fine structure and molecular cooling processes are extended to 10 \K.
In addition, many thermal processes exist, such as atomic line cooling from $\rm H$, $\rm He$ and metals, 
bremsstrahlung cooling, photoheating, photoelectric heating from dust grains, and cosmic ray heating (see table 1 of \citealt{Richings2014a}). 

A local approximation was used to shield the gas from the radiation field, in which a Sobolev-like approximation was used in relation to a density gradient to calculate a local shielding length. This shielding length is multiplied by the densities of individual species in order to determine their column densities, which are then used to attenuate photoionisation, photodissociation, and photoheating. It is assumed that the \HI~ cosmic ray ionization rate is $1.8 \times 10^{-16} \rm s^{-1}$ \citep{Indriolo2012}.

Given that our simulations only focus on the inner part of the circum-nuclear disk, we do not explicitly capture the stellar component of the host galaxy at larger radii (apart from through its gravitational potential). We therefore use a constant, uniform UV radiation field from Black 1987, measured in the Milky Way, for photoionisation and photoheating processes in \textsc{CHIMES}. 

\subsection{Star formation \& stellar feedback model}\label{sec:SFR}

This simulation is coupled with stellar feedback and ISM physics from FIRE-2 as described in \cite{Hopkins2018a}. There are a number of sources of stellar feedback in FIRE-2, such as radiation pressure (UV, optical, and IR, allowing for multiple scattering), stellar winds (O/B and AGB winds), supernovae (types Ia and II) and ionizing photons. 
The formation of stars is only possible in cold, molecular, and locally self-gravitating regions that have a number density above $\nh = 10^4 \, {\rm cm}^{-3}$ (as used in the most recent FIRE simulation studies e.g \citep{Torrey2020}).
The stellar evolution tracks, rates, yields, and kinetic luminosities are all tabulated based on stellar evolution models from STARBURST99 \citep{Leitherer1999} assuming the \textit{Kroupa} Initial Mass Function \citep[IMF;][]{Kroupa2002}, and  SNe Ia rates following \citet{Mannucci2006}. In the case of SNe II, Ia, and AGB winds, metal yields are taken from \citet{Nomoto2006}, \citet{Iwamoto1999}, and \citet{Izzard2004}, respectively. Each gas particle is tracked for the evolution of eleven species ($\rm H, He, C, N, O, Ne, Mg, Si, S, Ca,$ and $\rm Fe$). 

A wide range of observations related to galaxies has been successfully matched by FIRE, including the mass-metallicity relation and its evolution over redshift \citep{Ma2016} and the Kennicutt–Schmidt star formation law \citep{Orr2018}. Due to the high resolution, the star formation criteria, the cooling to low temperatures, and the multi-channel stellar feedback of FIRE, a reasonable ISM phase structure and giant molecular cloud (GMC) mass function has been obtained \citep{Benincasa2020}. Furthermore, these processes also result in a self-consistent development of galactic winds, which eject gas \citep{Muratov2015, Angles-alcazar2017} and metals \citep{Muratov2017, Hafen2019,Pandya2021} from galaxies to prevent excessive star formation and generate a plausible stellar-halo mass relation.
Additionally, the FIRE model is ideal for studying dust evolution due to its detailed treatment of the multiphase ISM and its ability to track the principal heavy elements that contribute to the formation of carbonaceous and silicate dust in gaseous form.

Moreover, we have local photo-ionization heating caused by young star particles. 
The number of photons per unit time produced by a star particle is determined from the STARBURST99 models. These are then used to ionise and heat the surrounding gas assuming the extent of a Stromgren sphere, starting with the nearest neighbour and moving outwards until the photon budget is exhausted. As it moves outward from the star, it ionizes each particle until it runs out of photons. Particles are flagged so they can not  cool below some minimum temperature $\sim10^4$ K for the rest of the timestep (and are heated to that temperature if already below it).
A long-range radiation pressure effect is also included in the simulation. 
It calculates each particle's luminosity and incorporates it into the gravity tree as a repulsive $1/r^2$ force, with the strength at each point based on the relevant absorption cross section integrated over three crude wavelength intervals in the ultraviolet, optical, and infrared - an approximation to a full spectrum. In this case, the minimum escape fraction from each star particle is determined by the fraction of emission in each wavelength interval. A noteworthy feature of CHIMES is that it enables coupling to radiation fields that vary in time and/or space rather than to non-equilibrium effects itself. It appears that our assumed radiation field is suitable for the solar neighborhood and probably too weak for our disk at the galactic center even if we ignore the AGN.
Nevertheless, in this study, the consideration of radiation feedback (including photo-ionization and dissociation) from the AGN is left for future studies and we focus only on mechanical feedback.
 
\subsection{Mechanical AGN feedback} \label{sec:AGN}

We use the black hole accretion model implemented by \citet{Hopkins2011}. In this approach, BH accretion is assumed to be determined by the gravitational torques 
\citep[see][for more details of accretion methods]{Hopkins2016}.
A captured particle adds its mass immediately to the BH accretion disk when it is gravitationally bound to the BH based on the assumption that the disks are supersonically turbulent, the accretion onto the BH can be written as follows: 
\begin{equation}
   \dot{\rm M}_{\rm BH} \sim \epsilon_{\rm acc} M_{\rm gas}(<R_0) \left( \frac{G M_{\rm total(<R_0)}}{R^3_0}\right)^{1/2},
\end{equation}
where $\epsilon_{\rm acc}$ is BH accretion factor (=1 in this simulation), $G$
is gravitational constant and $M_{\rm gas}$ and $M_{\rm total}$ are the gas and total mass in disc, respectively. 
The BH accretion luminosity scales with the accretion rate given the radiative efficiency, $L=\epsilon_{r}\,\dot{M}_{\rm BH}\,c^{2}$, where $\epsilon_{r}=0.1$ is assumed.  

In general, the mechanical feedback can be provided by synchrotron emitting jets\footnote{Jets are highly collimated beams of plasma that are launched from the vicinity of the black hole and can extend up to hundreds of kiloparsecs. They are usually detected in radio wavelengths and show high polarization due to synchrotron emission \citep{Silpa2022}.} and broad absorption lines (BAL) winds \footnote{Winds are less collimated outflows of gas that are driven by radiation pressure or magnetic fields from the accretion disk around the black hole. They are usually detected in optical or X-ray wavelengths and show broad emission lines due to Doppler shifts \citep{Silpa2022}}. Both jets and winds can have significant effects on the host galaxy and its surroundings, such as regulating star formation, transporting metals, heating or clearing the interstellar medium, etc.
Some studies have found a close association between radio structures and ionized gas morphology and kinematics in Seyfert galaxies, suggesting an interplay of jets, winds and emission-line gas in these sources. For example, some sources show an anti-correlation between polarized radio knots and [O III] emission, indicating that the radio emission is depolarized by the emission-line gas \citep[see e.g  Mrk 231, XID 2028 in][respectively]{Silpa2021,Cresci2023}.

It has been shown in numerous observational studies that slow jets are present in Seyfert galaxies (e.g. NGC~1068) through both the narrow line region (NLR) and radio outflow \citep[see][]{Axon1998, Roy2000, Gallimore1996a}. Other studies have found evidence of wind-driven outflows in Seyfert type 2 galaxies, such as NGC 1068, which show broad emission lines of ionized gas (such as [O III]) associated with the accretion disk \citep{Crenshaw2000}. In close proximity to the jet of NGC~1068, the morphology, kinematics, and, possibly, the ionization structure of the NLR correspond directly to the interaction with the radio outflow. Further, in accordance with \citet{Dugan2017}, jets may affect the pressure and density of the ambient medium differently from the effects of winds. 
The focus of our paper is solely on the observed ubiquitous AGN winds (not collimated), and we do not examine the differences between "jets" and "winds" in detail. 
In this study, AGNs exhibit blue-shifted BAL or troughs when their line of sight is intercepted by a high-speed outflow, likely originated from their accretion discs \citep[which is unresolved in the simulation; see][]{Hopkins2011}. Our scale disc is designed to shed light on the internal structure obscured by the AGN torus impacted by the outflow or wind.

Through the mechanical feedback process we implement in this simulation, wind mass and kinetic luminosity are continuously “injected” into the gas surrounding the SMBH where the outflow is isotropic \citep{Hopkins2016}.
It is assumed that some fraction of the photon momentum drives a wind at the resolution scale around the BH \citep{Murray1995}. Hence, the accreted gas is blown out as a wind with velocity $v_{\rm wind}$. The wind is defined by two parameters, the mass-loading of $\beta\equiv\dot{M}_{\rm wind}/\dot{M}_{\rm BH}$ and the velocity $v_{\rm wind}$ that equivalently relate to momentum-loading ($\dot{p}_{\rm wind}=\eta_{p}\,L/c$) and energy-loading ($\dot{E}_{\rm wind}=\eta_{E}\,L$) of the wind.
Following \cite{Hopkins2016}, the energy and momentum-loading are
\begin{align}\label{eq:momentum}
\eta_{P} &\equiv \frac{\dot{p}_{\rm wind}}{L/c} = \beta\,{\Bigl(}\frac{v_{\rm wind}}{\epsilon_{r}\,c}{\Bigr)}
\approx \beta\,{\Bigl(}\frac{v_{\rm wind}}{30,000\,{\rm km\,s^{-1}}}{\Bigr)}, \\ 
\eta_{E} &\equiv \frac{\dot{E}_{\rm wind}}{L} = \frac{\epsilon_{r}}{2}\,\frac{\eta_{P}^{2}}{\beta}
\approx0.05\,\beta\,{\Bigl(}\frac{v_{\rm wind}}{30,000\,{\rm km\,s^{-1}}}{\Bigr)}^{2}.
\end{align}
 
Our simulation has been carried out using different energy and momentum loading factors in five AGN models, including NoAGN \citep[with value for $\beta$ and $v_{\rm wind}$ in eqs. \ref{eq:momentum}, 3 based on observations and theoretical models; see for e.g.][]{Moe2009,Dunn2010,Hamann2011,Borguet2013}. All  the parameters are reported in Table~\ref{tab:AGN_table}.

\subsubsection{The model constraints}
In Figure \ref{fig:KS}, we show the observational consistency of our model in an analysis of the disk-averaged Kennicutt-Schmidt \citep[KS;][]{Kennicutt1998} relationships conducted at various times for models with AGN and NoAGN feedback. We have calculated the surface densities of \HI+\HH~ gas and star formation rates within 100 pc to compare to the observational range\footnote{The observational data is based upon a range of galaxies (not limited to galactic nuclei) with a scatter of $\pm 0.5$ dex as shown by the grey dotted line \citep{Narayanan2012} in Figure \ref{fig:KS}.} (see Figure \ref{fig:KS}). All models indicate that the CND moves along the KS 'main sequence' over time towards lower $\Sigma_{\rm gas}$ and lower $\Sigma_{\rm SFR}$. It is also evident from the NoAGN model that the SF integration frequency declines with time (see Figure \ref{fig:SFR}). Therefore, it is likely that a combination of either secular gas consumption by star formation in the CND, or the evacuation of gas outside the CND due to stellar feedback (SF-driven outflows) or the lack of replenishment of gas coming from outside the CND (i.e. no inflows) may explain the observed trends.

As shown in the left panel of Figure \ref{fig:SFR}, the star formation rate (SFR) varies with time for simulations incorporating mechanical feedback from the BH (AGN) and the model NoAGN. Star formation is suppressed by the AGN as compared to the NoAGN model.   
As shown in the figure, SFR ranges are comparable\footnote{Since the SFR measured in this paper extends to the specific radii of the disk, it cannot be accurately compared with the actual value of the SFR measured for the galaxies in observation. Our objective is only to demonstrate that the range of SFR of disk coincides with the observational range for the number of Seyfert galaxies.} to the observable reference for Seyfert galaxies \citep[see Tab. 1][for details and reference of SFR values]{Combes2019}. We are using global SFR that may correspond to kpc-disks to compare with the SFR values of our 200 pc model CNDs. Therefore, we restricted our analysis to time scales (3 - 8 Myr) when the SFR in disk is most consistent with Seyfert galaxies. 

The right panel of Figure \ref{fig:SFR} shows the evolution of the BH luminosity for various AGN models and the model NoAGN feedback. Our various AGN models yield an AGN whose luminosity ranges from $L \sim 10^{41} - 10^{46} erg\ s^{-1}$ across different time scales. These luminosities are much lower than those typically observed from the AGN at the centre of NGC~1068; however they are  consistent with those observed in most Seyfert galaxies \citep[see Table 7][for more details]{Combes2019}. Further, it is evident that there is a difference of $\sim$3 orders of magnitude between $ \rm L_{AGN}$ of NGC~1068 and the typical luminosity of the AGN models, but it is still below the range of the NoAGN model. The reason for this could be that firstly we did not include the bulge and stellar potential in this model and instead simply focused on the idealized disk model and secondly there are uncertainties regarding the BH accretion and feedback modules that were used in this study.
 \begin{figure}
	\centering
 \includegraphics[width=1.05\linewidth]{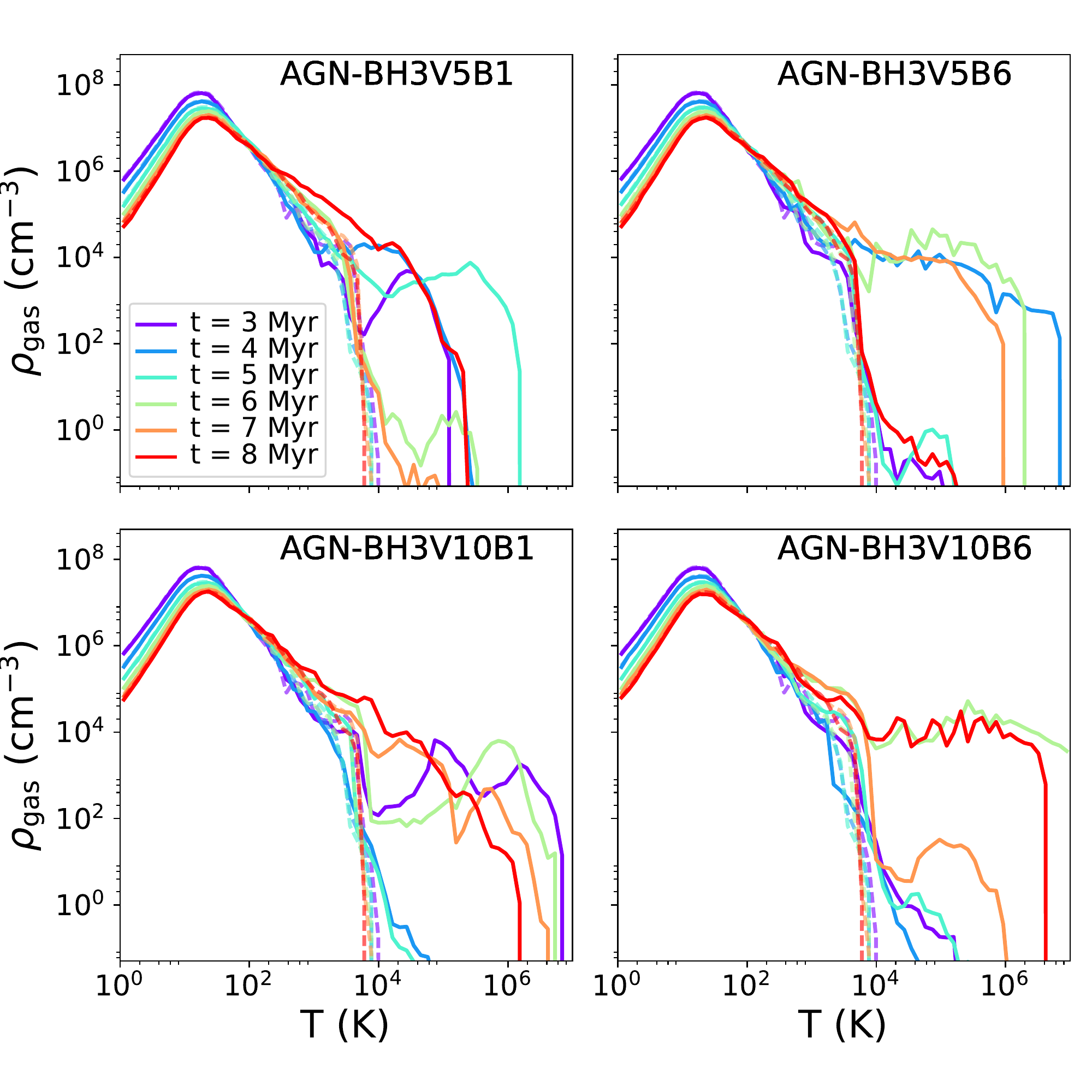}
	\caption{Density of gas versus temperature at disk scales for all AGN feedback models at a variety of times (colors). The model NoAGN is shown by color dashed lines in each panels. }
	\label{fig:rho_T}
\end{figure}

\begin{table}
        \centering
\caption{Parameters describing the simulations in the text including the Model name, $\eta_{p}$:Momentum-loading of AGN wind feedback($\dot{E}=\eta_{E}\,L$), $\eta_{E}$: Energy-loading of AGN wind feedback ($\dot{E}=\eta_{E}\,L$),$\beta$: AGN Mass-loading $\beta\equiv\dot{M}_{\rm wind}/\dot{M}_{\rm BH}$ (determined by $\eta_{p}$ \&\ $\eta_{E}$), $v_{\rm wind}$: AGN wind launching velocity at the simulation resolution (in ${\rm km\,s^{-1}}$; determined by $\eta_{p}$ \&\ $\eta_{E}$).}
        \label{tab:AGN_table}
        \large
\begin{tabular}{lcccc}
\hline
\hline
Model & $\eta_{p}$& $\eta_{E}$& $\beta$ &$v_{\rm wind}({\rm km\,s^{-1}})$\\
            \hline
AGN-BH3V5B1   &     0.1   &      0.001     &    1     &     5000     \\
AGN-BH3V5B6   &     1     &      0.008     &    6     &     5000     \\
AGN-BH3V10B1  &     0.3   &      0.005     &    1     &     10000    \\
AGN-BH3V10B6  &     2     &      0.03      &    6     &     10000    \\
NoAGN         &     0     &      0         &    0     &     0        \\ 
 \hline
\end{tabular}      
\end{table}

 \begin{figure}
	\centering
 \includegraphics[width=1.05\linewidth]{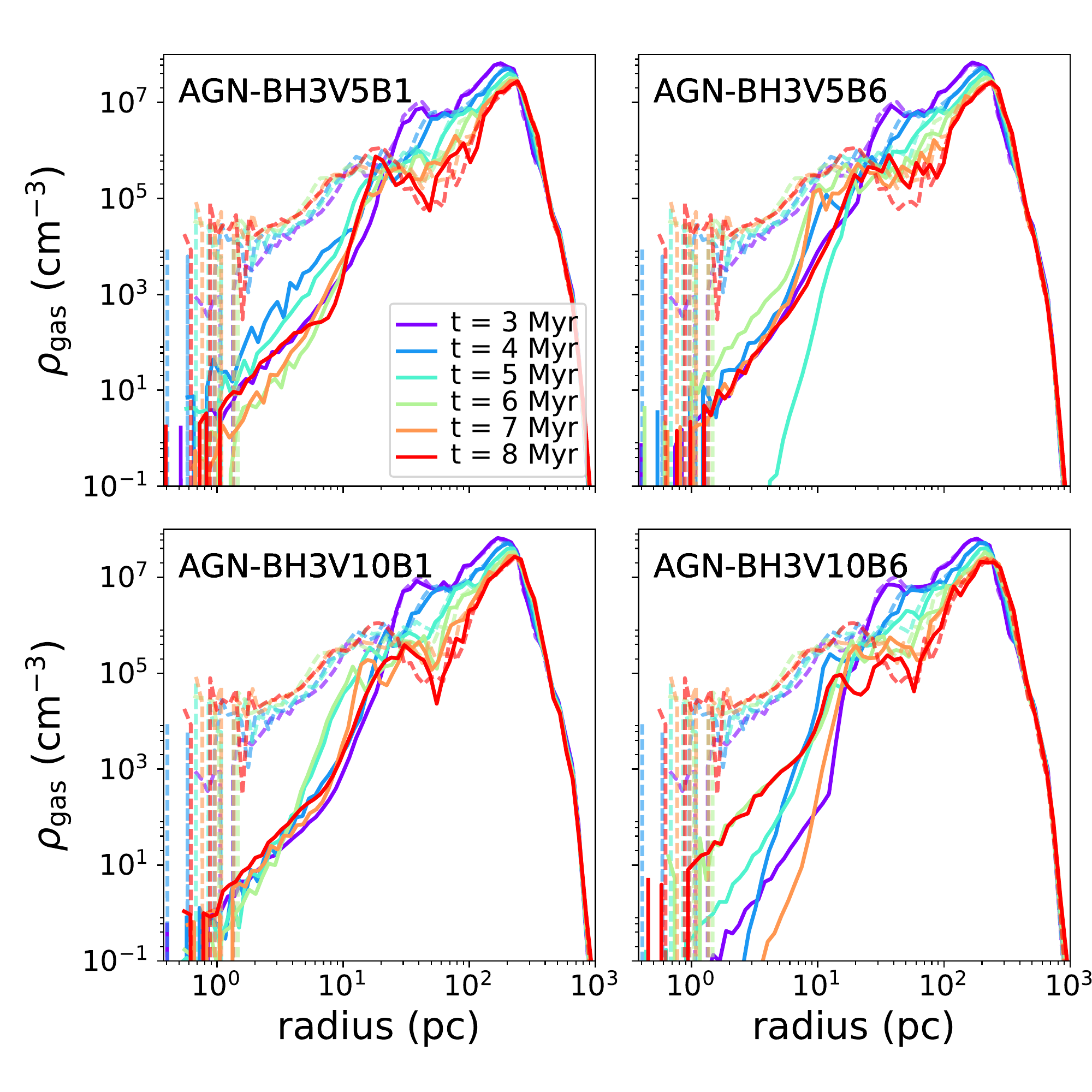}
	\caption{Gas density profiles at disk scales for all AGN feedback models at a variety of times (colors). The model NoAGN is shown by color dashed lines in each panel.}
	\label{fig:rho_p}
\end{figure}

 \begin{figure}
	\centering
 \includegraphics[width=1.05\linewidth]{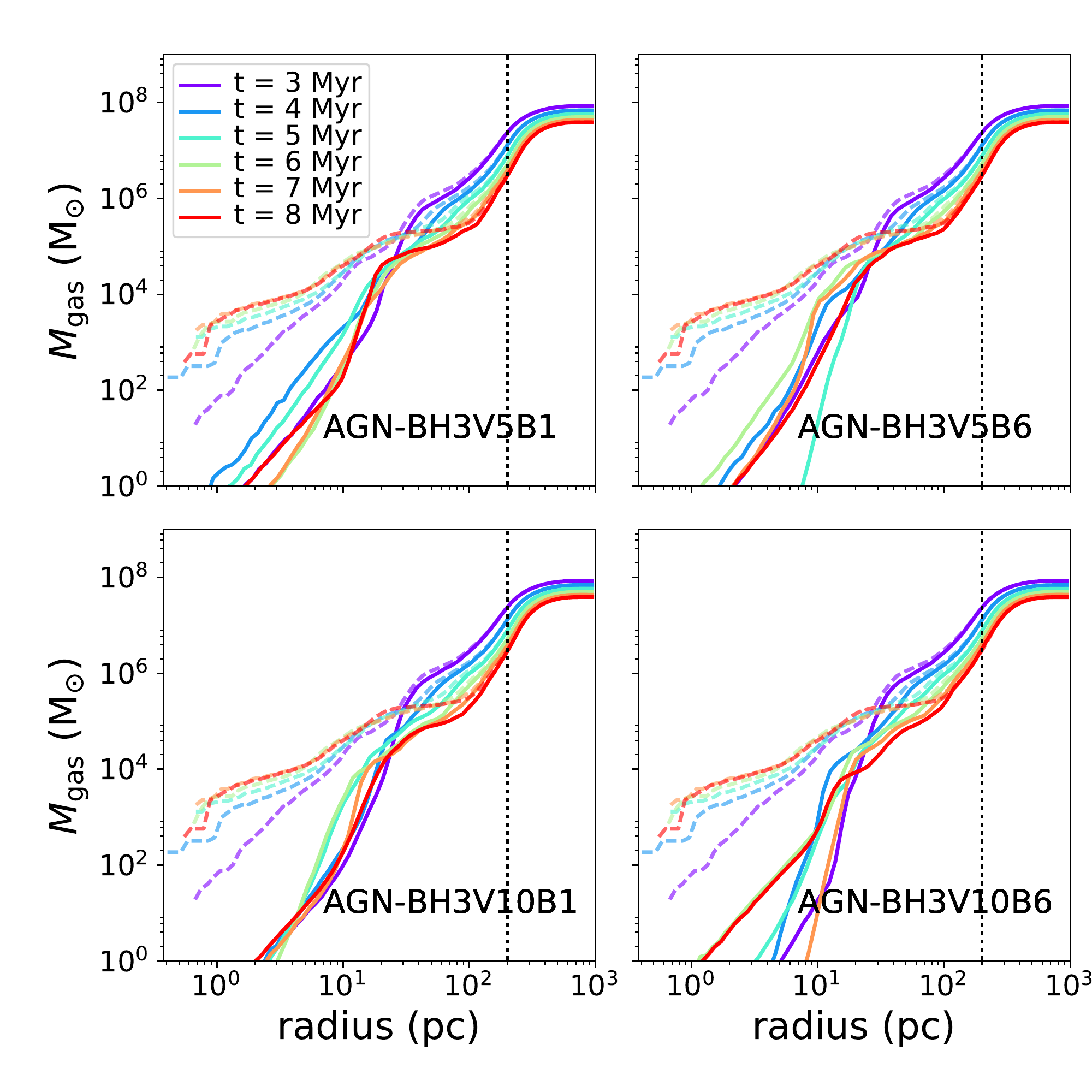}
	\caption{Gas mass profiles at disk scales for all AGN feedback models at a variety of times (colors). The model NoAGN is shown by color dashed lines in each panel. Grey Dotted line shows the CND scale of disk $\sim$200 pc corresponding to $\sim10^7 M_{\odot}$ gas mass. }
	\label{fig:Mass_p}
\end{figure}

 \begin{figure*}
	\centering
 \includegraphics[width=1.07\linewidth]{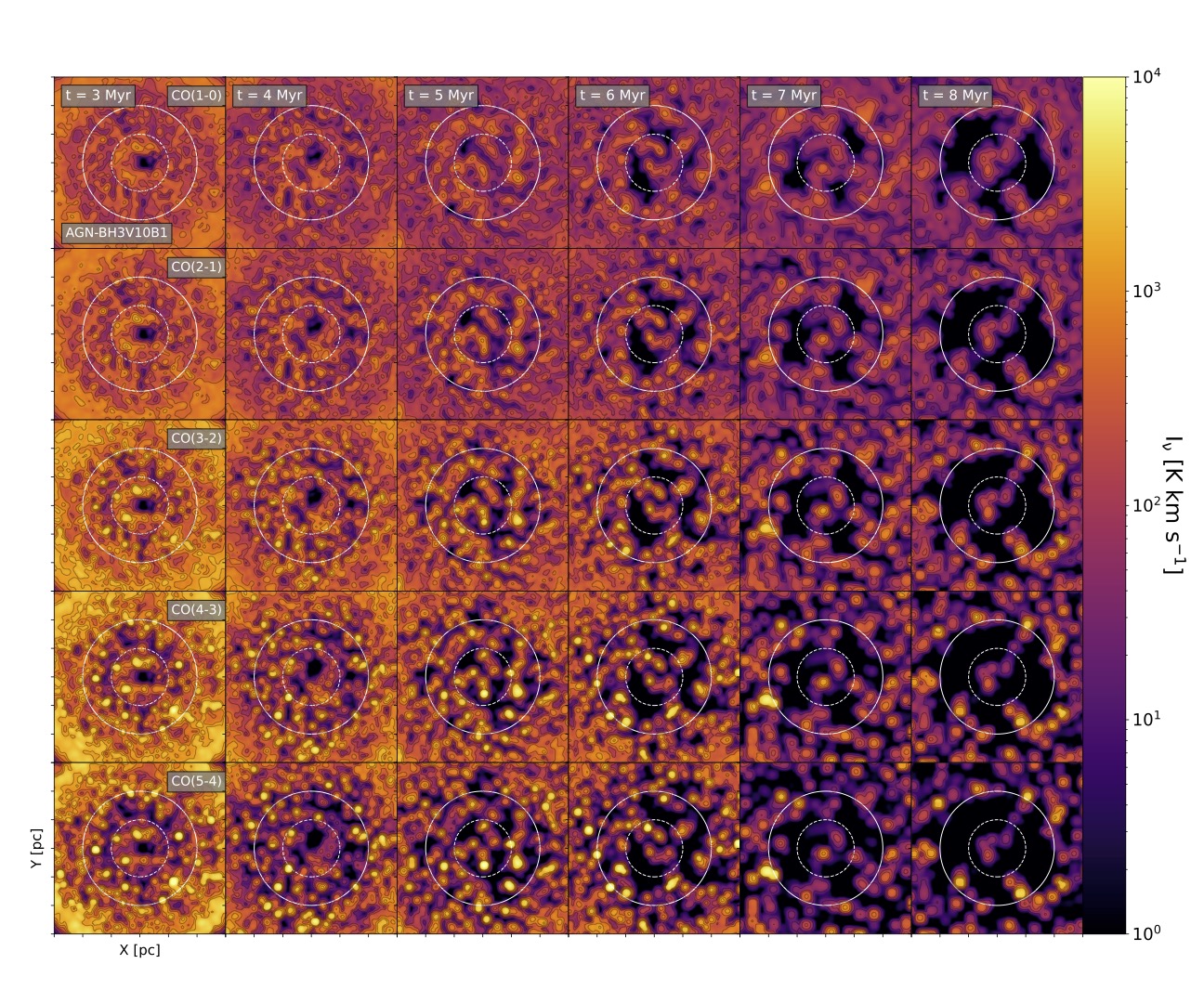}
	\caption{
	Integrated CO line intensity maps (Moment-0) obtained with RADMC-3D radiative transfer code on simulation data for different emission lines (J=1-0, 2-1, 3-2, 4-3, 5-4; from top to bottom ) and times (3-8 Myr from left to right) for a model with mechanical AGN feedback (BH3V10B1). Throughout the images, the beam size is 0.06"$\times$0.06" (corresponding to 4.2 pc at the comoving radial distance of 14 Mpc for NGC 1068) with position angle equal to zero degrees. In each panel, the image size is a central 150 pc of disk radius. The contours range from 25$\sigma$ to 1600$\sigma$, and follow each other by factor 2 multiplication. The circles represent the central 50 and 100 pc radius of the disk. 
	}
	\label{fig:Moment0}
\end{figure*}

\subsection{Radiative transfer calculations}\label{sec:RT} 

The maps of emission lines are generated through the post-processing of simulation snapshots using version 0.40 of the publicly available Monte Carlo radiative transfer code  \textsc{radmc-3d} \citep{Dullemond2012}, utilizing the abundances of ions and molecules calculated during the simulations with the \textsc{chimes} chemistry module. 
Due to the fact that \textsc{radmc-3d} is a grid-based code, we first project the gas particles from the simulations onto an Adaptive Mesh Refinement (AMR) grid, which is refined so that no cell contains more than 10 particles. 
To avoid unphysical effects that may arise from mixing particles with very different properties into the same cell, we weight the contribution of each particle according to its abundance when we project the temperatures and velocities of the gas onto the grid. We are only modelling emission and attenuation from the CND itself, with the caveat that this does not include attenuation from the host galaxy since the CO lines are at relatively long wavelengths where dust is not really a problem. However, we note that there could also be contamination from CO lines coming from the host galaxy itself, which is not captured by this simulation, as we are only modelling the signal from the inner CND.

In all cases, the level populations are calculated by \textsc{radmc-3d}, using an approximate non-LTE approach based on the Local Velocity Gradient (LVG) method. In order to generate the synthetic CO emission maps in the full face-on and inclined (41$^\circ$) views, we made use of molecular data (energy levels, transition probabilities, collisional excitation rates) from the Leiden (\textsc{lamda}) database \footnote{\url{https://home.strw.leidenuniv.nl/\~moldata/}} \citep{Schoier2005}. 
The CO line images are produced with a passband ranging from -100 km/s to +100 km/s and 400 wavelength points evenly distributed around the line's center. In this study, we generated the synthetic CO emission maps for CO(1-0), CO(2-1), CO(3-2), CO(4-3) and CO(5-4). Using again NGC 1068 as an example, we use the distance of 14 Mpc (so that 70 pc equals 1" in each moment map). 

\section{Analysis and Results} \label{sec:results1}
\subsection{Temperature and density  projections}

A face-on view (x-y projection) of the volume density and temperature of the gas at various times (3-8 Myr with a step of 1 Myr per evolution) are shown in Figures \ref{fig:Map_density} and \ref{fig:Map_Temp} for various AGN feedback models.
As shown in  figure \ref{fig:Map_density}, a CND scale disk of gas is centered over the central BH, illustrating the volume density evolution at different times. A broad range of volume gas densities ($1$ to $10^5\ cm^{-3}$) is visible from the center to the outer disk region, which is in line with what we expect in the CND at the seyfert-2 type galaxies \citep[e.g NGC~1068 see;][]{Scourfield2020}, with the lowest median density seen since the start of the simulation, at 8 Myr.
As time progresses, the high density regions begin to form at radii smaller than 100~pc from the center and expand more towards the end of the simulation. At a given time, the AGN models show different densities within the central r=100~pc  of the disk compared with the NoAGN model. This can be attributed to the difference in energy and momentum loading factors in the models.   
Unlike the AGN models, the gas in the NoAGN model concentrates in the centre rather than dispersing over time by wind or accretion. 
As both the velocity and the mass-loading factor increase in the AGN-BH3V10B6 model, the gas is largely removed from the disk. 

In Figure \ref{fig:Map_Temp} the gas temperature fluctuates between  $10\ K$  (the minimum temperature in the simulation) to $10^6 K$ and increases over the course of the expansion at the center due to the BH activity. As the simulation proceeds, the AGN central heating begins to be efficient at 3 Myr and impacts the environment  in a different manner from the NoAGN model. Again, the heating of the gas by the AGN is attributed to the energy and momentum loading factors in the AGN models, where at the disk center the gas temperature rises above $10^6 K$, especially as the wind velocity increases.  In the NoAGN model, the temperature is generally below $10^4 K$. 

At fixed wind velocity, we run simulations using high and low mass-loading, wherein the BH accretion rate decreased and increased by about an order of magnitude, respectively. As can be seen from the right panel of Figure \ref{fig:SFR}, the AGN luminosity reflects this.
Accordingly, a higher mass-loading produces initially stronger outflows at higher temperatures in Figure \ref{fig:Map_Temp}, leading to slightly lower volume density sightlines in Figure \ref{fig:Map_density}.
The average AGN luminosity (BH accretion rate) remains relatively similar (right panel of Figure \ref{fig:SFR}) at low mass-loading of the SMBH (BH3V10B1,BH3V5B1). This results in a similar effect on the volume density as shown in  Figure \ref{fig:Map_density}, but a slightly greater effect on the temperature (see Figure \ref{fig:Map_Temp}). As wind momentum is conserved in this limit, the absolute wind speed (at fixed mass-loading) makes little qualitative difference to the dynamics but the higher velocity produces more shock-heated high-temperature gas. 
In the presence of high wind velocity and differing mass-loading (BH3V10B1,BH3B10B6), we see interesting effects that result in higher AGN luminosity (BH accretion rates) as shown in right panel of Figure \ref{fig:SFR}. This results in a strong effect on the volume density and temperature (see Figures \ref{fig:Map_density} and \ref{fig:Map_Temp}).
A greater mass-loading results in a greater net wind momentum injection and outflow rate.

\subsubsection{Temperature-Density relation, Gas density and mass profiles}
   
Figure \ref{fig:rho_T} shows the evolution of temperature-density for the disk over time for various AGN models compared to the model without AGN feedback. The gas distribution is governed by the temperature-density phase space, and it is evident that the presence of the AGN affects the distributions. Since the energy and momentum loading factors differ among the various AGN models, the density increases at different rates as the temperature increases, and it is significantly higher in the model with both high velocity and mass-loadings for the wind. 
For most AGN models, the mechanical feedback results in higher temperature and lead to clumpiness  due to the compression of the gas.
 Also, Figure \ref{fig:rho_p} shows the density profile of gas at a given radii of the disk over time for various AGN models compared to the model without AGN feedback. As can be seen, there is a lower density of gas for all AGN models, particularly at small scale radii ($<$50 pc) in proximity of SMBH due to AGN heating.
Both AGN and NoAGN  models tend to end up producing a ring morphology. Although AGN-driven outflow can be the underlying cause of the creation of a conspicuous nuclear gas deficit due to the removal of gas it is likely not the sole cause. As such, the NoAGN model ending up in a ring morphology may be due to either a stellar feedback effect or gas consumption by the SF.  
Observationally, the ring morphology in the CND of NGC~1068 is the result of the combined effect of gravity torques at the  Inner Lindblad Resonance (ILR) of the bar, which pile up gas at this resonance, and the AGN wind which removes gas inside the ILR \citep[e.g. see][]{Garcia-Burillo2019}.

In Figure \ref{fig:Mass_p}, the mass profile of the gas for the disk is shown for same time scales. In general the models are consistent  with the measured mass of the CND and tori in the disk of Seyfert galaxies \citep[i.e $M_{\rm tori}/M_{\rm Disk}\sim$ 0.1; see][and references therein]{Combes2019} where the gas mass accumulation in the central 200 pc of disk and tori ($\sim$ 50 pc) are around $10^6\ M_{\odot}$ and $10^5\ M_{\odot}$, respectively.
Less gases present in the central region of the disk for AGN models and at some point the gas pushes out to the outer region when the mass-loading factor increases (i.e BH3V5B6 and BH3V10B6).   
As a consequence, models that incorporate AGN feedback are expected to compresses and push out the gas resulting in the clumpy (high density) regions.

\subsection{CO Molecular lines emission}

Figure \ref{fig:Moment0} shows the CO integrated intensity map of the CND (moment-0) for times between 3 and 8 Myr for AGN model BH3V10B1. In particular, we show the CO(1-0), CO(2-1), CO(3-2), CO(4-3) and CO(5-4) transitions  since these are commonly observed lines in extragalactic environments. As time goes on, the molecular gas is scattered outward from the central region of the CND ($>$100 pc).This can be attributed to the AGN feedback on the gas in the center, which is likely to destroy or dynamically push the molecular gas. We also note that the pressure of the central hot bubble drives an outflow that compresses the ambient ISM at larger radii of the disk, which would tend to promote CO formation \citep{Zubovas2014,Richings2021}. Indeed, this is in agreement with Figures \ref{fig:rho_T},\ref{fig:rho_p}, which illustrate how the presence of the AGN affects the distribution of density and temperature at a given radius. We see that the corresponding CO formation requires more time if feedback is not present. According to the comparison of gas density in the AGN and NoAGN runs, since AGN feedback compresses the gas, such higher gas densities result in a shorter time-scale for the formation of CO.

Finally we see that there is a deficit of molecular gas in the central region of the CND, especially at the scale of r=50~pc. Throughout the disk, on average, the intensity in clumpy regions appears to be higher for the higher CO $J$ lines.

 \begin{figure*}
	\centering
	\includegraphics[width=1.07\linewidth]{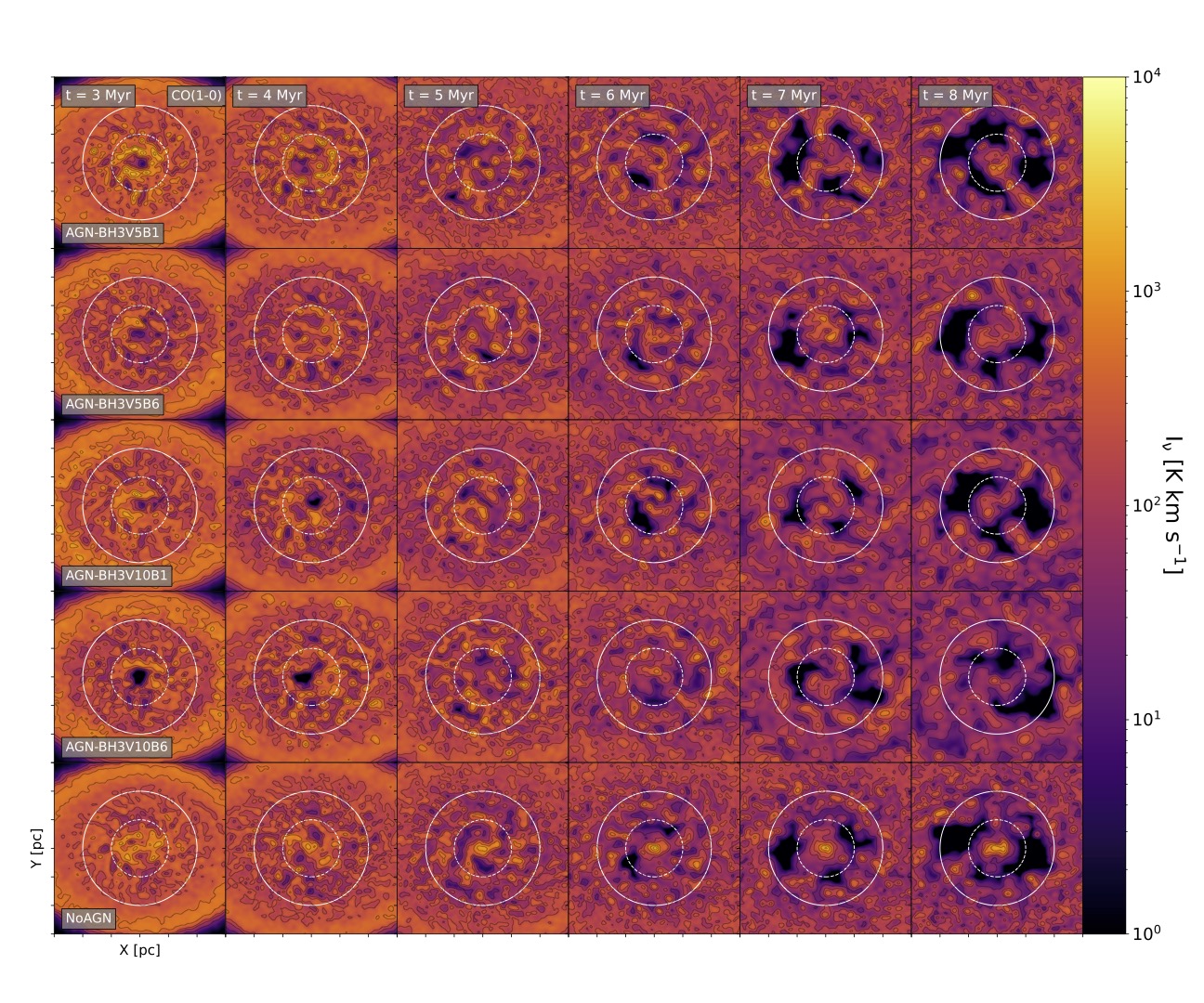}
	\caption{ A map of the CO(1-0) line integrated intensity (Moment-0) for various AGN models and NoAGN feedback obtained using the RADMC-3D radiative transfer code with an inclination of 41 degrees, in correspondence to that of NGC 1068 for times between 3 and 8 Myr from left to right). Throughout the images, the beam size is 0.06"$\times$0.06" (corresponding to 4.2 pc at the comoving radial distance of 14 Mpc for NGC 1068). The contours range from 25$\sigma$ to 1600$\sigma$, and follow each other by factor 2 multiplication. The circles represent the central 50 and 100 pc of the disk. }
\label{fig:Moment0_NGC1068}
\end{figure*}

 \begin{figure*}
	\centering
	\includegraphics[width=1.07\linewidth]{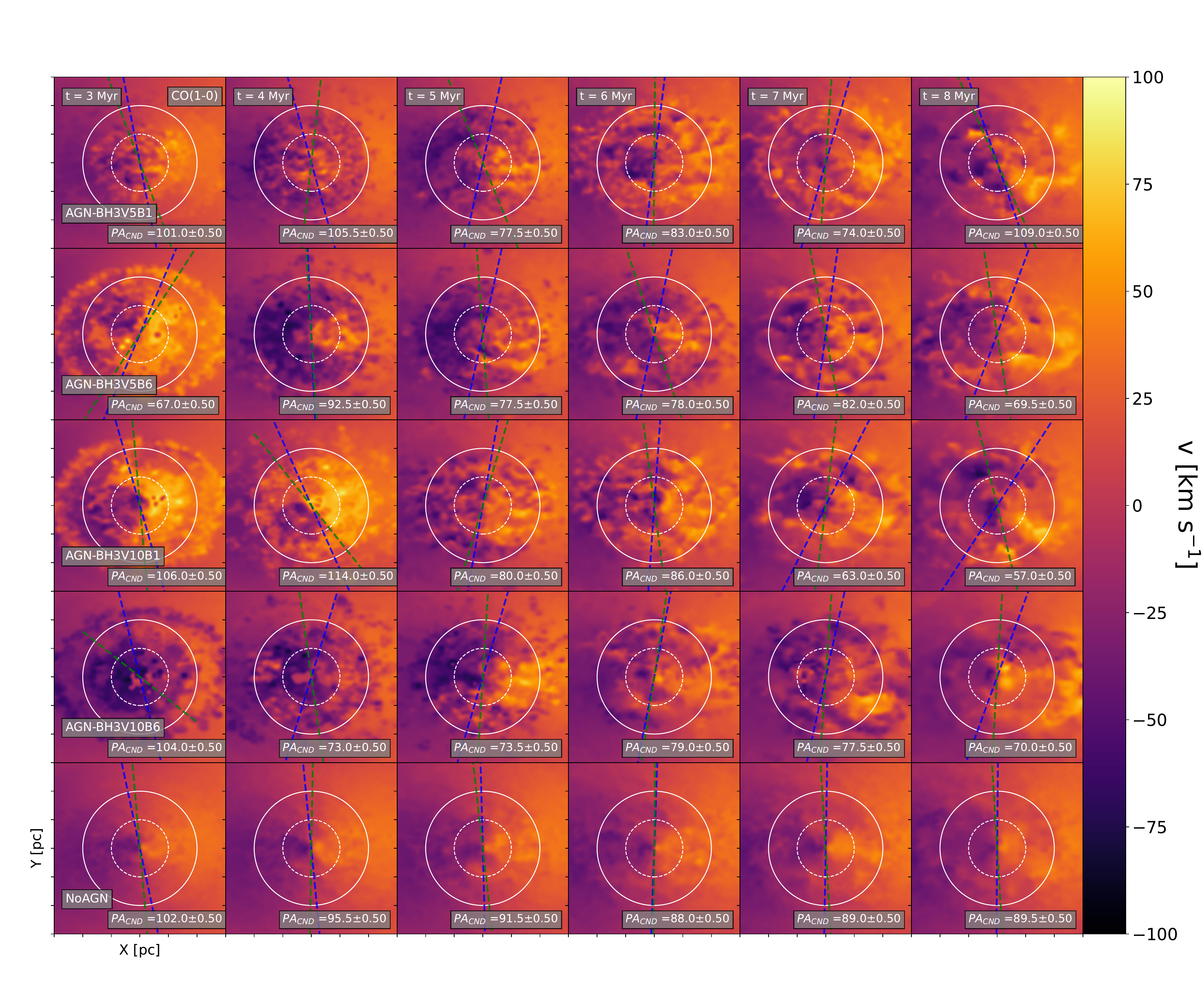}
	\caption{A map of the CO(1-0) line velocity map (Momment-1) for various AGN model and NoAGN feedback obtained using the RADMC-3D radiative transfer code with an inclination of 41 degree, in correspondence to that of NGC 1068, for times between 3 and 8 Myr from left to right). The circles represents the central r=500~pc(dotted) and 100~pc of the disk. Throughout the images, the beam size is 0.06"$\times$0.06" (corresponding to 4.2 pc at the comoving radial distance of 14 Mpc for NGC 1068). The blue and green dashed lines in the panels indicate the minor kinematic position angles ($\rm PA\pm90$) of CND in the central  r=50 pc and 100 pc  scale of the disk, respectively. A value of $PA_{CND}$ corresponds to the minor kinematic position angle for r=100~pc. }
	\label{fig:Moment1_NGC1068}
\end{figure*}

 \begin{figure}
	\centering
	\includegraphics[width=1.\linewidth]{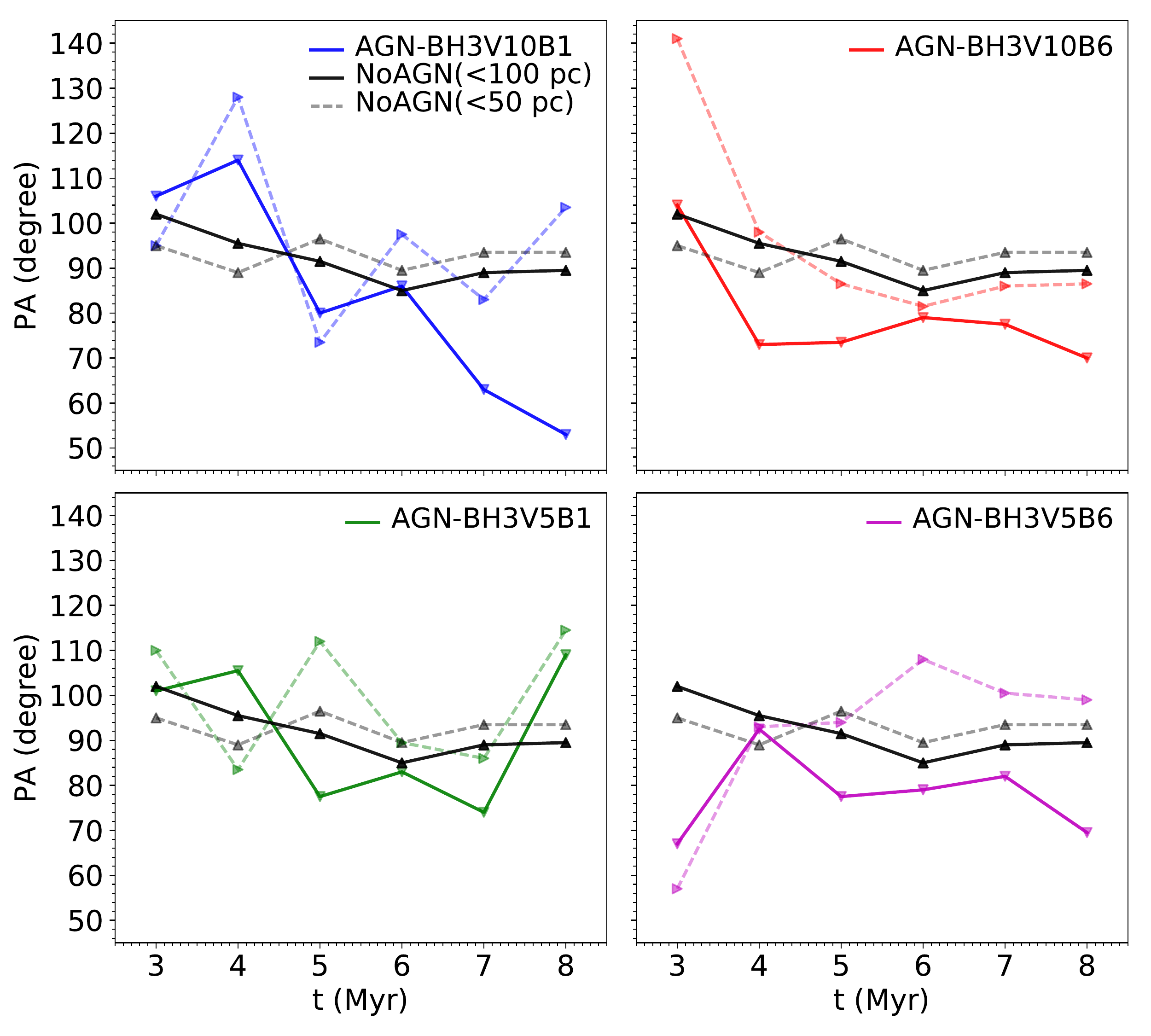}
	\caption{Kinematic position angles of CO(1-0) moment-1 maps within the central 50 (dashed lines) and 100 (solid lines) pc of the disk for AGN models with different energy and momentum loading factors and the model NoAGN (black -lines) measured at times between 3-8 Myr.}
	\label{fig:PA_angle}
\end{figure}

 \begin{figure*}
	\centering
	\includegraphics[width=1.07\linewidth]{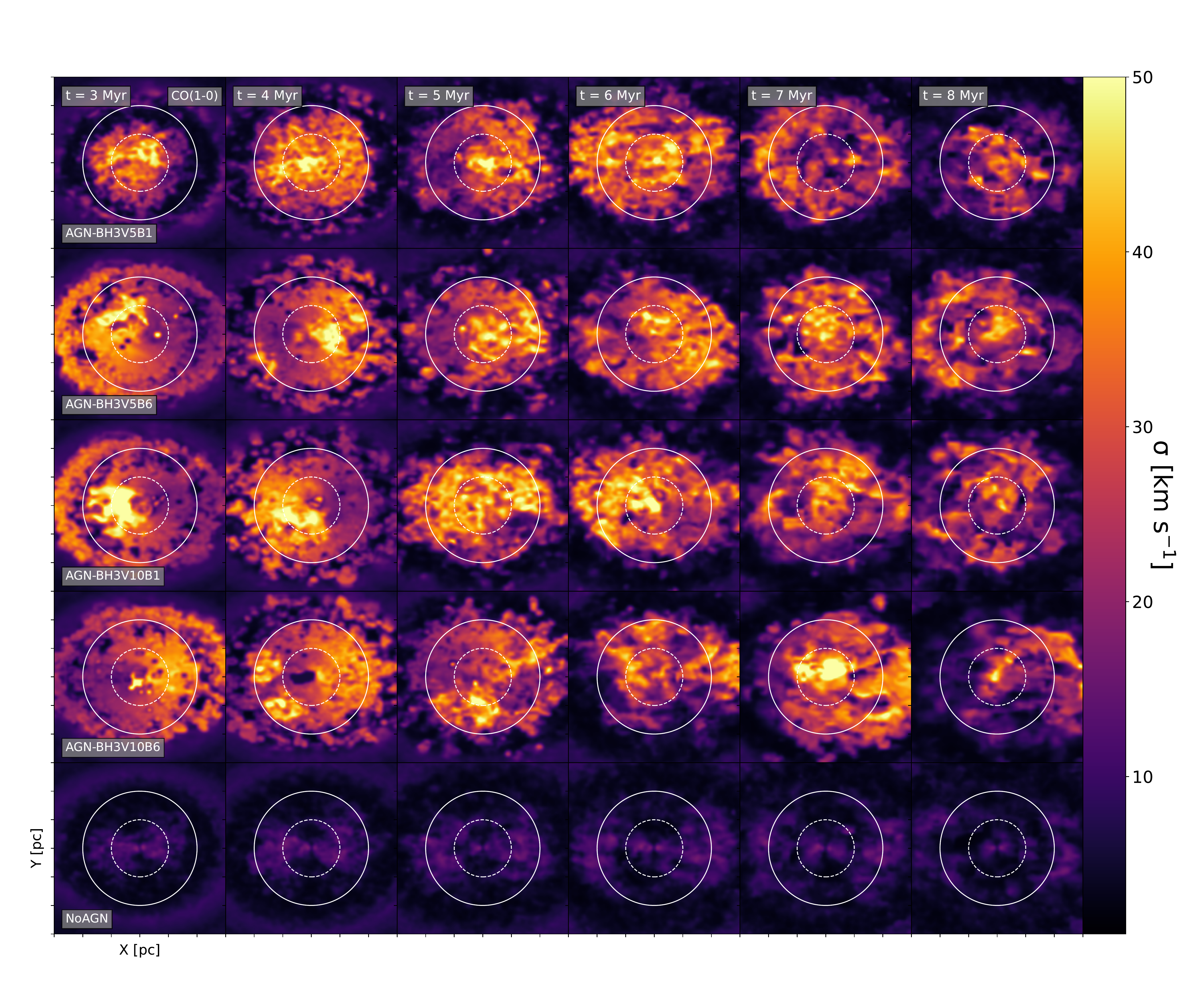}
	\caption{ A map of the CO(1-0) line dispersion ($\sigma$; Moment-2) for various AGN models and NoAGN feedback obtained using the RADMC-3D radiative transfer code with an inclination of 41 degree, in correspondence to that of NGC 1068, for times between 3 and 8 Myr from left to right). The circle represents the central r=100~pc of the disk.}
	\label{fig:Moment2_NGC1068}
\end{figure*}

 \begin{figure*}
	\centering
         \includegraphics[width=1.07\linewidth]{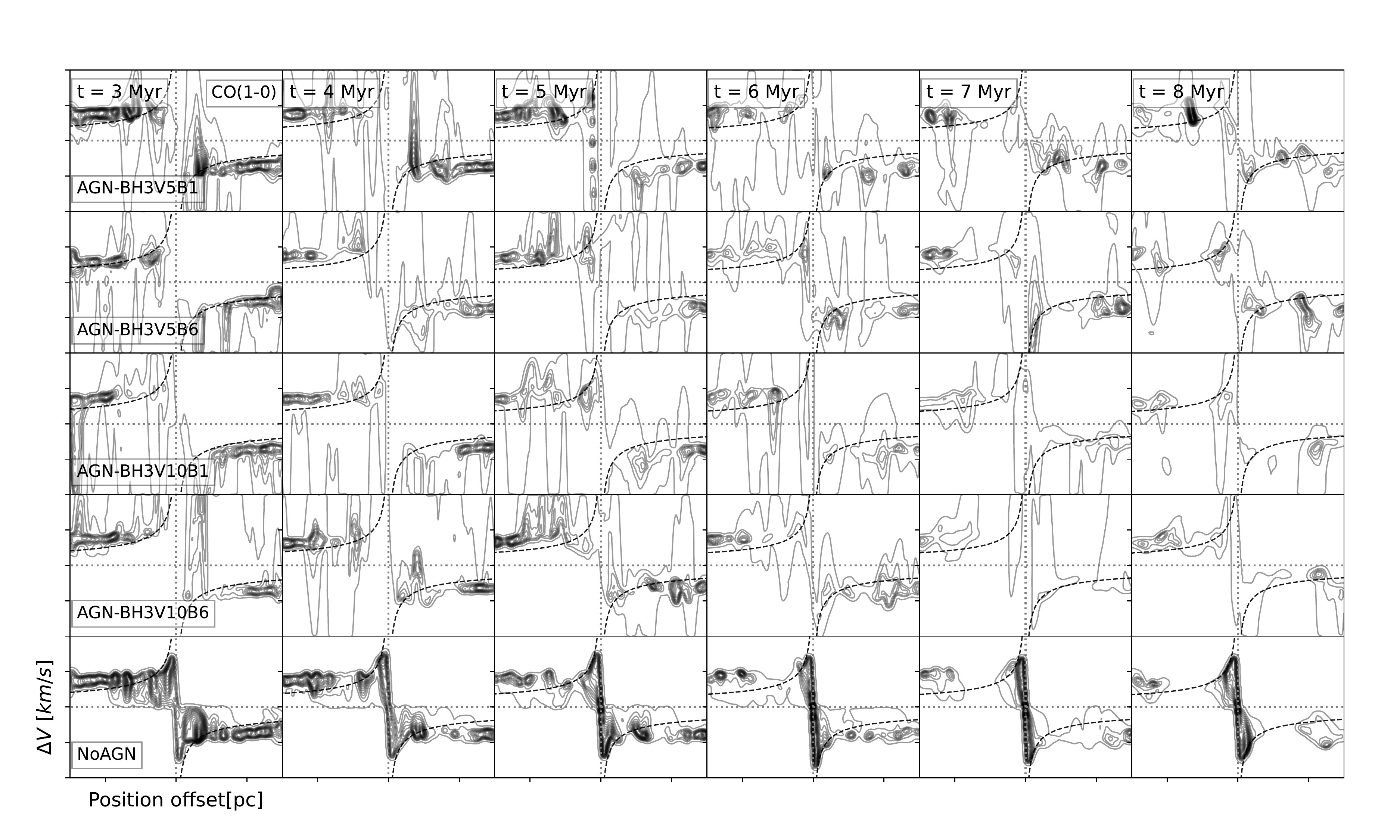}
  	\caption{ A contours of the CO(1-0) line position-velocity over the major axis of inclined disk $PA_{CND}$ for various AGN models and the model NoAGN feedback for time scale between 3 and 8 Myrs (left to right). Every panel shows a position offset between -150 and 150 pc and a velocity between -100 and 100 km/s. The dashed-black curves show the Keplerian rotation curve fit, $v_{rot}\propto r^{\alpha}$ for mass within the central r=100~pc scale of disk where here $\alpha = -0.5$.}
	\label{fig:PV_NGC1068}
\end{figure*}

\subsection{Comparisons with the CO emission in NGC~1068}
In this section we attempt a qualitative comparison of our simulations with the observed CO emission in NGC 1068.  We note that in our simulations  the cosmic ray ionization  rate is assumed to be $1.8 \times 10^{16}$ which is the galactic value for diffuse clouds \citep[e.g; Milky Way][]{Indriolo2012}. This may indeed not be correct as the cosmic ray ionization rate in AGN-dominated galaxies may be higher \citep[e.g;][]{Aladro2013,Viti2014,Scourfield2020}. However, to a first order approximation, the CO intensities should not be too affected by a modest increase in cosmic ray ionization rate. Moreover, a 10$^{-16}$ s$^{-1}$ cosmic ray ionization rate is an upper limit for our Galaxy, associated with diffuse gas: in galactic dense molecular cloud this rate may be lower by one order of magnitude \citep{Li2003}. Nevertheless, future simulations of NGC~1068 should consider spanning a range of cosmic ray ionization rates. In the following, we present inclined moments maps for various AGN models and the model NoAGN for different times.  

\subsubsection{CO(1-0) integrated intensity map; Moment-0}
Figure \ref{fig:Moment0_NGC1068} shows the CO(1-0) integrated intensity with an inclination of 41$^\circ$ \citep[which corresponds to NGC 1068's inclination as firstly reported in][]{Bland-Hawthorn1997} for various AGN models and times between 3 to 8 Myr. Every panel has been overplotted with contours ranging from 25$\sigma$ to 1600$\sigma$ with factor 2 multiplication. 
As in  Figure \ref{fig:Moment0}, the integrated intensity maps in the top rows show less contrast in the ring morphology of the CND, but the gas appears to pile up in a ring and moves out at least the region between r=50~pc and r=100~pc in NoAGN model. 
Moreover, in the cases with AGN, CO  could form more efficiently given sufficient time due to the presence of AGN feedback - this is in contrast with the NoAGN model. Therefore the enhanced CO is seen in the cases with AGN feedback at the central r=50~pc scale.

We perform a qualitative comparison with the CO observations at a $\sim$ 40 pc resolution in \citet{Viti2014} (see their Table 3) -  where the CO(1-0) integrated intensity ranges from 200 to 600 $\rm K\ kms^{-1}$, with the highest values around the East and West knots. 
Concentrating on the inner 50 pc of the CND, we also see that the observed range of CO(1-0) intensities are easily matched in most regions up to 6 Myr, with a possible best match for AGN-BH3V5B6, a model with a low velocity wind and high mass-loading factor. Observational studies have been conducted to demonstrate that the ISM gas around SMBH is dramatically affected by high-velocity winds ($\gtrsim$ 2000 $\rm km/s$) with momentum fluxes of $0.1-1L/c$ \citep[e.g.][]{Cimatti2013,Harrison2014,Zakamska2016}. 

\subsubsection{The kinematics of the CND}
Figure \ref{fig:Moment1_NGC1068} shows the velocity maps (moment-1) for various AGN models and the model NoAGN feedback, which are derived from the CO(1-0) emission line. 
Within the central r=100~pc, the apparent kinematic major axis of the CND is approximately oriented along the west/east direction, coincidentally, and the velocity ranges from -100 to 100 km/s. 
The position angle (PA) of the 2D gas rotation pixels that pass the quality cut (after removing the outlier pixels, $\gtrsim$ 3$\sigma$, using a $\sigma$-clipping approach) for the velocity field is measured using the \textsc{pafit} code in pypi\footnote{https://pypi.org/project/pafit/}  described in \cite{Krajnovic2006}. A kinematic PA measurement assumes that the center of the map is at (0,0) and measured counterclockwise, with 0 degrees being North in the maps. 
As shown by the PA fit, there is misalignment since the PA changes both with time and radius for AGN models by more than 20 degrees\footnote{The exact values of these cut-offs do not impact the main conclusions of this paper in most cases and are simply derived from observations for the offset between stellar and gas kinematics PAs. \citep[see e.g][]{Lagos2015,Bryant2019}}  in comparison to NoAGN. 
It is interesting to note that the inflow phase dominates the dynamics at the earlier timeframes ($\sim$3, 4 Myr) when both mass loading and wind velocity are high (BH3V10B6). 
It can be seen that there is a PA offset at different radii in various AGN models. A larger offset  in PA occurs when strong winds cause gas expulsion by AGN. Regardless of whether outflows occurs inside or outside the plane, it will tilt the PA by 90 degrees. The PA minor can also be tilted by inflows, but with the opposite sign.  In some AGN models, the tilt is also changing sign as a function of time, which implies that either the orientation of the outflows from the plane changes compared to that of the large-scale disk, or that inflows from the plane reverse the sign of the tilt at certain stages of the simulation.

Figure \ref{fig:PA_angle} shows the evolution of kinematic PAs of the CO(1-0) velocity map on scales of r=100~pc and r=50~pc as a function of time. There are several panels in the figure that represent offsets for PA over time or at different scales around the BH compared with the NoAGN model which differ slightly in both cases (i.e.  aligned PA).  
A large proportion of such PA offsets occur within the central r=50~pc of the disk surrounding the SMBH in low mass-loading wind models (BH3V10B1, BH3V5B1), where the change in PA occurs within a shorter timeframe. There is a significant early offset (at 3-4 Myr) in models with a high mass-loading factor (BH3V10B6, BH3V5B6), followed by the steady PA trend to the later times. Moreover, different scale radii of AGN models with low velocity winds (BH3V5B6) exhibit a higher PA offset.
The different radii reveal similar trends for all models except for the model with high wind velocity (i.e., BH3V10B1), which exhibits a large PA offset at a later time (t$\sim$8 Myr) compared to an earlier times.

Our evaluation of the effect of energy and momentum loading factors for various AGN feedback models on the PA offset at different radii reveals that model BH3V5V1 exhibits re-alignment when comparing the first and last timeframes, whereas the other models require more time to align. 
As the mass-loading increases, both BH3V10B6 and BH3V5B6 exhibit a different early orientation than their low mass-loading counterparts, then remain stable over time. NGC 1068 observations indicate that the counter rotational motion of the disk \citep[i.e PA offset across different scale radii of more than 150 degrees, e.g. see][]{Raouf2021} is dynamically unstable \citep{Garcia-Burillo2019}. As a result, it is evident that the mechanical AGN feedback could contribute to explaining the stability of PA in the CND scale disk around the SMBH through the loading of momentum and energy. 

\subsubsection{Velocity Dispersion}
Figure \ref{fig:Moment2_NGC1068} shows  measurements of the velocity dispersion, $\sigma$ (moment-2),  made in the inclined CND for the various AGN models and the model NoAGN feedback. AGN feedback increases $\sigma$ from 10 to 50 km/s. 
It is noteworthy that the 2D pattern of $\sigma$ changes significantly with time. If the enhanced velocity dispersion is due to the onset of outflows, which alter the rotation kinematics of the disk, then this behavior implies that outflows are 'rapidly' evolving over time in the AGN models on timescales of 1 Myr.
In some regions of the CND, the molecular gas exhibits a higher $\sigma$, which is a result of the superposition of different velocity components associated with rotation, as well as an underlying pronounced outflow pattern. 
In general, the pattern varies with  time and increases of $\sigma$ towards the central region. 
Further, the superposition of different velocity components associated with rotation is not present in the NoAGN simulation because there is no pattern associated with the outflow. Comparing different AGN models, there is a large scatter of change in the superposition. The model with high velocity and high mass-loading factor (BH3V10B5) has a very high $\sigma$ in the central r=50~pc of the disk at $t\sim7$ Myr. 

\subsubsection{P-V diagram}
In Figure \ref{fig:PV_NGC1068}, the time evolution of position-velocity (p-v) contours, obtained along the average fitted major axis of the nuclear disk at $PA_{CND}$, provides further insight into the kinematics of the molecular gas disk for various AGN and NoAGN models (top to bottom). Every panel shows a position offset between -150 and 150 pc and a velocity between -100 and 100 $\rm km/s$. 
In all models, most of the emission occurs in the top left and bottom right quadrants of the diagram. A Keplerian rotation curve ($v_{rot}\propto r^{\alpha}$ with $\alpha = -0.5$) fitting the central 150 pc disk mass is overplotted on the contour plots. 
The velocities are somewhat consistent with the Keplerian rotation curve, especially at earlier times ($\sim$3 Myr); however, the scatter is large. Note that at large radii, all of the rotation curves appear to be super-Keplerian.
When the models are evaluated at 5 Myr, the models with higher velocity and different mass loading factors (BH3V10B6, BH3V10B1) exhibit a smaller scatter than those with low velocity winds.  Further, all models are more or less consistent with Keplerian rotation at 8 Myr. On the other hand, the disk in the NoAGN simulation roughly follows the Keplerian rotation at all times (with the best fit at earlier times), and it displays a smaller degree of scatter than the various AGN models.

\section{Discussion and Summary}
\label{subsec:sum}
We present a hydrodynamic model of the gas disk of an AGN dominated galaxy that incorporates direct radiative cooling, non-equilibrium chemistry, and mechanical feedback by the central supermassive black hole. The different emission line intensity maps are evaluated through the use of RADMC-3D, a post-processing radiative transfer code. We compare the model with different mechanical AGN feedback energy and momentum loading factors with the model without AGN (NoAGN). We constrain our models by using Seyfert-2 galaxies reported in the observational studies of \citet{Combes2019}. In the kinematic approach our simulations are then qualitatively compared to observations of the AGN-dominated galaxies (e.g NGC~1068). Across the simulated circumnuclear disk, the volume gas densities, temperatures and gas mass profiles are in agreement with observations of  NGC 1068. Additionally, the focus is also on the times ($\sim$3-8 Myr) at which the AGN luminosity and the SFR range can be compared to a number of Seyfert 2 galaxies reported in table 1. \citet{Combes2019}.
 The main results are as follows:
 
	\begin{enumerate}
	  \item 
         The model including an AGN promotes the formation of CO in the clumpy regions around the supermassive blackhole. This can be seen in the CO intensity maps, 
         as well as the corresponding changes in gas density and temperature, when compared with the model without AGN feedback. As the distance from the central supermassive blackhole increases ($>$50 pc), on average, there appears to be an increase in the intensity of higher-J CO lines.
         At later time, in the central region of the CND ($r<$100 pc), there is a deficit of molecular gas consistent with ALMA observations \citet{Garcia-Burillo2019}.

         \item
        Through the different AGN models the CO(1-0) inclined intensity maps show different features at the central r=100~pc and 50~pc scales of the disk. 
        The observed CO(1-0) intensities can easily be matched in most regions up to 6 Myr within the inner 50 pc of the CND, with a possible best fit for a model with high mass-loading factors. This is consistent with the dramatic effects of AGN on the ISM around SMBHs, observationally. 
	\item
        It is evident that the mechanical AGN feedback could contribute to explaining the alignment stability of the PA in the CND scale disk around the SMBH through the loading of momentum and energy. The observations indicate that the counter rotational motion of the disk, where the PA offset exceeds 150 degrees, is dynamically unstable \citep{Garcia-Burillo2019}.
        The wind causes the disk to be out of alignment, with the PA offset being high for a longer period of time on the 100 pc scale of disk (i.e continue to be misaligned). The disk appears to re-align within a short period of time when the wind velocity is lower. As the mass-loading is increased, the disk smooths out following an initial large offset. It is interesting to note that the inflow phase dominates the dynamics at earlier timeframes ($\sim$3, 4 Myr) when mass loading and wind velocity are high. It is clear that the model NoAGN does not show a significant PA offset at any time or radius. 
	
       \item  
        For the models with AGN feedback, molecular gas has a higher velocity dispersion, $\sigma$, in certain regions of the CND, which is due to the superposition of velocity components that can be attributed to the outflow.
        By contrast, the NoAGN model does not exhibit the enhanced velocity dispersion due to the outflow.    

       \item
        A position-velocity diagram (p-v) indicates that the velocities are broadly consistent with a Keplerian rotation curve, particularly at earlier times ($\sim$3 Myr) and larger radii; however, the rotation velocity scatter is large. 
        At 5 Myr, the models with higher velocity and different mass loading exhibit a smaller scatter than those with low velocity winds. Moreover, all models are more or less consistent with Keplerian rotation at 8 Myr. Furthermore, in the absence of AGN feedback, the disk clearly follows the Keplerian rotation across all time scales. It exhibits a lower degree of scatter in the rotation velocity than in the presence of AGN feedback. 
        	        
    \end{enumerate}

    In conclusion, this study suggests that the SMBH at the center of the disk, as well as AGN feedback, may have a significant impact on the dynamics of molecular gas, particularly in the central 50 and 100 pc of the disk. During the evolution of the CND, the dynamics of the molecular gas may be unstable due to outflows from  the central regions and the rotation of the gas. Intensity maps of CO lines and the dynamics are affected by the energy and momentum loading factors of mechanical AGN feedback. Changing the wind velocity and mass-loading factors produces different dynamics in the central 50 and 100 pc of the CND. In general the disk has a misaligned PA in the central r=100~pc for models with AGN feedback. There seems to be a connection between the AGN wind velocity and the disk PA offset. Accordingly, it is clear that the kinematic properties of the molecular gas change as time progresses, in some cases leading to a large PA offset (misalignment) every $\sim$ 2 Myr from the onset of AGN feedback.
    It appears that the rotation of the disk is affected by the presence of outflows when comparing the position-velocity diagram for the models with and without feedback from the AGN. An indication of this can be seen in the non-keplerian and keplerian rotation curves in the evolution of the disk in the models with and without AGN respectively. 
    In view of the complexity of modeling a particular galaxy, our objective is to evaluate the effect of mechanical AGN feedback on the gas surrounding the SMBH. In this regard, we do not compare our model directly to NGC~1068 galaxy, but rather used it as a typical AGN-dominated galaxy.

    Future work will focus on the analysis of different molecular line ratios with the aim of tracing and characterizing individual energetic processes in AGN-dominated galaxies. A comparison with a run assuming equilibrium chemistry and different ionization rates is also planned. In addition, we aim to conduct numerical simulations that incorporate the gravitational potential of the stellar disk and bulge components as part of the subsequent future work addressing AGN feedback.

\section*{Acknowledgments} \label{sec:acknow}
  We thank the anonymous referee for their comments that have enriched this work. MR thanks Philip Hopkins for his helpful conversations during the conference in Sweden. An Advanced Research Grant from the European Union (833460) supported this study as part of the  MOlecules as Probes of the Physics of EXternal (Moppex) galaxies project. SGB acknowledges support from the research project PID2019-106027GA-C44 of the Spanish Ministerio de Ciencia e Innovaci\'on.

\section*{Data availability}
Data from the simulations is based on the output from publicly available GIZMO \citep{Hopkins2015} code at the following repository: \href{https://bitbucket.org/phopkins/gizmo-public/src/master/}{https://bitbucket.org/phopkins/gizmo-public/src/master/}  generated on the ICs described in section \ref{sec:simulation}. \textit{The data underlying this article will be shared on reasonable request to the corresponding author}.

\appendix

\bibliographystyle{mnras}
\bibliography{bibliography.bib}

\begin{thebibliography}{}
\makeatletter
\relax
\def\mn@urlcharsother{\let\do\@makeother \do\$\do\&\do\#\do\^\do\_\do\%\do\~}
\def\mn@doi{\begingroup\mn@urlcharsother \@ifnextchar [ {\mn@doi@}
  {\mn@doi@[]}}
\def\mn@doi@[#1]#2{\def\@tempa{#1}\ifx\@tempa\@empty \href
  {http://dx.doi.org/#2} {doi:#2}\else \href {http://dx.doi.org/#2} {#1}\fi
  \endgroup}
\def\mn@eprint#1#2{\mn@eprint@#1:#2::\@nil}
\def\mn@eprint@arXiv#1{\href {http://arxiv.org/abs/#1} {{\tt arXiv:#1}}}
\def\mn@eprint@dblp#1{\href {http://dblp.uni-trier.de/rec/bibtex/#1.xml}
  {dblp:#1}}
\def\mn@eprint@#1:#2:#3:#4\@nil{\def\@tempa {#1}\def\@tempb {#2}\def\@tempc
  {#3}\ifx \@tempc \@empty \let \@tempc \@tempb \let \@tempb \@tempa \fi \ifx
  \@tempb \@empty \def\@tempb {arXiv}\fi \@ifundefined
  {mn@eprint@\@tempb}{\@tempb:\@tempc}{\expandafter \expandafter \csname
  mn@eprint@\@tempb\endcsname \expandafter{\@tempc}}}

\bibitem[\protect\citeauthoryear{{Aalto}, {Garcia-Burillo}, {Muller},
  {Winters}, {van der Werf}, {Henkel}, {Costagliola}  \& {Neri}}{{Aalto}
  et~al.}{2012}]{Aalto2012}
{Aalto} S.,  {Garcia-Burillo} S.,  {Muller} S.,  {Winters} J.~M.,  {van der
  Werf} P.,  {Henkel} C.,  {Costagliola} F.,   {Neri} R.,  2012, \mn@doi [\aap]
  {10.1051/0004-6361/201117919}, \href
  {https://ui.adsabs.harvard.edu/abs/2012A&A...537A..44A} {537, A44}

\bibitem[\protect\citeauthoryear{{Aladro} et~al.,}{{Aladro}
  et~al.}{2013}]{Aladro2013}
{Aladro} R.,  et~al., 2013, \mn@doi [\aap] {10.1051/0004-6361/201220131}, \href
  {https://ui.adsabs.harvard.edu/abs/2013A&A...549A..39A} {549, A39}

\bibitem[\protect\citeauthoryear{{Angl{\'e}s-Alc{\'a}zar},
  {Faucher-Gigu{\`e}re}, Kere{\v s}, Hopkins, Quataert  \&
  Murray}{{Angl{\'e}s-Alc{\'a}zar} et~al.}{2017}]{Angles-alcazar2017}
{Angl{\'e}s-Alc{\'a}zar} D.,  {Faucher-Gigu{\`e}re} C.-A.,  Kere{\v s} D.,
  Hopkins P.~F.,  Quataert E.,   Murray N.,  2017, \mn@doi [\mnras]
  {10.1093/mnras/stx1517}, 470, 4698

\bibitem[\protect\citeauthoryear{{Axon}, {Marconi}, {Capetti}, {Macchetto},
  {Schreier}  \& {Robinson}}{{Axon} et~al.}{1998}]{Axon1998}
{Axon} D.~J.,  {Marconi} A.,  {Capetti} A.,  {Macchetto} F.~D.,  {Schreier} E.,
    {Robinson} A.,  1998, \mn@doi [\apjl] {10.1086/311249}, \href
  {https://ui.adsabs.harvard.edu/abs/1998ApJ...496L..75A} {496, L75}

\bibitem[\protect\citeauthoryear{{Barbosa}, {Storchi-Bergmann}, {McGregor},
  {Vale}  \& {Rogemar Riffel}}{{Barbosa} et~al.}{2014}]{Barbosa2014}
{Barbosa} F.~K.~B.,  {Storchi-Bergmann} T.,  {McGregor} P.,  {Vale} T.~B.,
  {Rogemar Riffel} A.,  2014, \mn@doi [\mnras] {10.1093/mnras/stu1637}, \href
  {https://ui.adsabs.harvard.edu/abs/2014MNRAS.445.2353B} {445, 2353}

\bibitem[\protect\citeauthoryear{Benincasa et~al.,}{Benincasa
  et~al.}{2020}]{Benincasa2020}
Benincasa S.~M.,  et~al., 2020, \mn@doi [\mnras] {10.1093/mnras/staa2116}, 497,
  3993

\bibitem[\protect\citeauthoryear{{Bland-Hawthorn}, {Gallimore}, {Tacconi},
  {Brinks}, {Baum}, {Antonucci}  \& {Cecil}}{{Bland-Hawthorn}
  et~al.}{1997}]{Bland-Hawthorn1997}
{Bland-Hawthorn} J.,  {Gallimore} J.~F.,  {Tacconi} L.~J.,  {Brinks} E.,
  {Baum} S.~A.,  {Antonucci} R.~R.~J.,   {Cecil} G.~N.,  1997, \mn@doi [\apss]
  {10.1023/A:1000567831370}, \href
  {https://ui.adsabs.harvard.edu/abs/1997Ap&SS.248....9B} {248, 9}

\bibitem[\protect\citeauthoryear{{Booth} \& {Schaye}}{{Booth} \&
  {Schaye}}{2009}]{Booth2009}
{Booth} C.~M.,  {Schaye} J.,  2009, \mn@doi [\mnras]
  {10.1111/j.1365-2966.2009.15043.x}, \href
  {https://ui.adsabs.harvard.edu/abs/2009MNRAS.398...53B} {398, 53}

\bibitem[\protect\citeauthoryear{{Borguet}, {Arav}, {Edmonds}, {Chamberlain}
  \& {Benn}}{{Borguet} et~al.}{2013}]{Borguet2013}
{Borguet} B. C.~J.,  {Arav} N.,  {Edmonds} D.,  {Chamberlain} C.,   {Benn} C.,
  2013, \mn@doi [\apj] {10.1088/0004-637X/762/1/49}, \href
  {https://ui.adsabs.harvard.edu/abs/2013ApJ...762...49B} {762, 49}

\bibitem[\protect\citeauthoryear{{Bower}, {Schaye}, {Frenk}, {Theuns},
  {Schaller}, {Crain}  \& {McAlpine}}{{Bower} et~al.}{2017}]{Bower2017}
{Bower} R.~G.,  {Schaye} J.,  {Frenk} C.~S.,  {Theuns} T.,  {Schaller} M.,
  {Crain} R.~A.,   {McAlpine} S.,  2017, \mn@doi [\mnras]
  {10.1093/mnras/stw2735}, \href
  {https://ui.adsabs.harvard.edu/abs/2017MNRAS.465...32B} {465, 32}

\bibitem[\protect\citeauthoryear{{Bryant} et~al.,}{{Bryant}
  et~al.}{2019}]{Bryant2019}
{Bryant} J.~J.,  et~al., 2019, \mn@doi [\mnras] {10.1093/mnras/sty3122}, \href
  {https://ui.adsabs.harvard.edu/abs/2019MNRAS.483..458B} {483, 458}

\bibitem[\protect\citeauthoryear{{Cimatti} et~al.,}{{Cimatti}
  et~al.}{2013}]{Cimatti2013}
{Cimatti} A.,  et~al., 2013, \mn@doi [\apjl] {10.1088/2041-8205/779/1/L13},
  \href {https://ui.adsabs.harvard.edu/abs/2013ApJ...779L..13C} {779, L13}

\bibitem[\protect\citeauthoryear{{Combes} et~al.,}{{Combes}
  et~al.}{2019}]{Combes2019}
{Combes} F.,  et~al., 2019, \mn@doi [\aap] {10.1051/0004-6361/201834560}, \href
  {https://ui.adsabs.harvard.edu/abs/2019A&A...623A..79C} {623, A79}

\bibitem[\protect\citeauthoryear{{Crenshaw} \& {Kraemer}}{{Crenshaw} \&
  {Kraemer}}{2000}]{Crenshaw2000}
{Crenshaw} D.~M.,  {Kraemer} S.~B.,  2000, \mn@doi [\apj] {10.1086/308570},
  \href {https://ui.adsabs.harvard.edu/abs/2000ApJ...532..247C} {532, 247}

\bibitem[\protect\citeauthoryear{{Cresci} et~al.,}{{Cresci}
  et~al.}{2023}]{Cresci2023}
{Cresci} G.,  et~al., 2023, \mn@doi [\aap] {10.1051/0004-6361/202346001}, \href
  {https://ui.adsabs.harvard.edu/abs/2023A&A...672A.128C} {672, A128}

\bibitem[\protect\citeauthoryear{{Croton} et~al.,}{{Croton}
  et~al.}{2006}]{Croton2006}
{Croton} D.~J.,  et~al., 2006, \mn@doi [\mnras]
  {10.1111/j.1365-2966.2005.09675.x}, \href
  {https://ui.adsabs.harvard.edu/abs/2006MNRAS.365...11C} {365, 11}

\bibitem[\protect\citeauthoryear{{Das}, {Crenshaw}, {Kraemer}  \& {Deo}}{{Das}
  et~al.}{2006}]{Das2006}
{Das} V.,  {Crenshaw} D.~M.,  {Kraemer} S.~B.,   {Deo} R.~P.,  2006, \mn@doi
  [\aj] {10.1086/504899}, \href
  {https://ui.adsabs.harvard.edu/abs/2006AJ....132..620D} {132, 620}

\bibitem[\protect\citeauthoryear{{Dav{\'e}}, {Angl{\'e}s-Alc{\'a}zar},
  {Narayanan}, {Li}, {Rafieferantsoa}  \& {Appleby}}{{Dav{\'e}}
  et~al.}{2019}]{Dave2019}
{Dav{\'e}} R.,  {Angl{\'e}s-Alc{\'a}zar} D.,  {Narayanan} D.,  {Li} Q.,
  {Rafieferantsoa} M.~H.,   {Appleby} S.,  2019, \mn@doi [\mnras]
  {10.1093/mnras/stz937}, \href
  {https://ui.adsabs.harvard.edu/abs/2019MNRAS.486.2827D} {486, 2827}

\bibitem[\protect\citeauthoryear{{Di Matteo}, {Springel}  \& {Hernquist}}{{Di
  Matteo} et~al.}{2005}]{DiMatteo2005}
{Di Matteo} T.,  {Springel} V.,   {Hernquist} L.,  2005, \mn@doi [\nat]
  {10.1038/nature03335}, \href
  {https://ui.adsabs.harvard.edu/abs/2005Natur.433..604D} {433, 604}

\bibitem[\protect\citeauthoryear{{Di Matteo}, {Colberg}, {Springel},
  {Hernquist}  \& {Sijacki}}{{Di Matteo} et~al.}{2008}]{DiMatteo2008}
{Di Matteo} T.,  {Colberg} J.,  {Springel} V.,  {Hernquist} L.,   {Sijacki} D.,
   2008, \mn@doi [\apj] {10.1086/524921}, \href
  {https://ui.adsabs.harvard.edu/abs/2008ApJ...676...33D} {676, 33}

\bibitem[\protect\citeauthoryear{{Dubois}, {Gavazzi}, {Peirani}  \&
  {Silk}}{{Dubois} et~al.}{2013}]{Dubois2013}
{Dubois} Y.,  {Gavazzi} R.,  {Peirani} S.,   {Silk} J.,  2013, \mn@doi [\mnras]
  {10.1093/mnras/stt997}, \href
  {https://ui.adsabs.harvard.edu/abs/2013MNRAS.433.3297D} {433, 3297}

\bibitem[\protect\citeauthoryear{{Dugan}, {Gaibler}  \& {Silk}}{{Dugan}
  et~al.}{2017}]{Dugan2017}
{Dugan} Z.,  {Gaibler} V.,   {Silk} J.,  2017, \mn@doi [\apj]
  {10.3847/1538-4357/aa7566}, \href
  {https://ui.adsabs.harvard.edu/abs/2017ApJ...844...37D} {844, 37}

\bibitem[\protect\citeauthoryear{{Dullemond}, {Juhasz}, {Pohl}, {Sereshti},
  {Shetty}, {Peters}, {Commercon}  \& {Flock}}{{Dullemond}
  et~al.}{2012}]{Dullemond2012}
{Dullemond} C.~P.,  {Juhasz} A.,  {Pohl} A.,  {Sereshti} F.,  {Shetty} R.,
  {Peters} T.,  {Commercon} B.,   {Flock} M.,  2012, {RADMC-3D: A multi-purpose
  radiative transfer tool}, Astrophysics Source Code Library, record
  ascl:1202.015 (\mn@eprint {ascl} {1202.015})

\bibitem[\protect\citeauthoryear{{Dunn} et~al.,}{{Dunn}
  et~al.}{2010}]{Dunn2010}
{Dunn} J.~P.,  et~al., 2010, \mn@doi [\apj] {10.1088/0004-637X/709/2/611},
  \href {https://ui.adsabs.harvard.edu/abs/2010ApJ...709..611D} {709, 611}

\bibitem[\protect\citeauthoryear{{Frias Castillo} et~al.,}{{Frias Castillo}
  et~al.}{2022}]{FriasCastillo2022}
{Frias Castillo} M.,  et~al., 2022, \mn@doi [\apj] {10.3847/1538-4357/ac6105},
  \href {https://ui.adsabs.harvard.edu/abs/2022ApJ...930...35F} {930, 35}

\bibitem[\protect\citeauthoryear{{Gallimore}, {Baum}, {O'Dea}  \&
  {Pedlar}}{{Gallimore} et~al.}{1996}]{Gallimore1996a}
{Gallimore} J.~F.,  {Baum} S.~A.,  {O'Dea} C.~P.,   {Pedlar} A.,  1996, \mn@doi
  [\apj] {10.1086/176798}, \href
  {https://ui.adsabs.harvard.edu/abs/1996ApJ...458..136G} {458, 136}

\bibitem[\protect\citeauthoryear{{Gallimore}, {Henkel}, {Baum}, {Glass},
  {Claussen}, {Prieto}  \& {Von Kap-herr}}{{Gallimore}
  et~al.}{2001}]{Gallimore2001}
{Gallimore} J.~F.,  {Henkel} C.,  {Baum} S.~A.,  {Glass} I.~S.,  {Claussen}
  M.~J.,  {Prieto} M.~A.,   {Von Kap-herr} A.,  2001, \mn@doi [\apj]
  {10.1086/321616}, \href
  {https://ui.adsabs.harvard.edu/abs/2001ApJ...556..694G} {556, 694}

\bibitem[\protect\citeauthoryear{{Gallimore} et~al.,}{{Gallimore}
  et~al.}{2016}]{Gallimore2016}
{Gallimore} J.~F.,  et~al., 2016, \mn@doi [\apjl] {10.3847/2041-8205/829/1/L7},
  \href {https://ui.adsabs.harvard.edu/abs/2016ApJ...829L...7G} {829, L7}

\bibitem[\protect\citeauthoryear{{Garc{\'\i}a-Burillo}
  et~al.,}{{Garc{\'\i}a-Burillo} et~al.}{2010}]{Garcia-Burillo2010}
{Garc{\'\i}a-Burillo} S.,  et~al., 2010, \mn@doi [\aap]
  {10.1051/0004-6361/201014539}, \href
  {https://ui.adsabs.harvard.edu/abs/2010A&A...519A...2G} {519, A2}

\bibitem[\protect\citeauthoryear{{Garc{\'\i}a-Burillo}
  et~al.,}{{Garc{\'\i}a-Burillo} et~al.}{2014}]{Garcia-Burillo2014}
{Garc{\'\i}a-Burillo} S.,  et~al., 2014, \mn@doi [\aap]
  {10.1051/0004-6361/201423843}, \href
  {https://ui.adsabs.harvard.edu/abs/2014A&A...567A.125G} {567, A125}

\bibitem[\protect\citeauthoryear{{Garc{\'\i}a-Burillo}
  et~al.,}{{Garc{\'\i}a-Burillo} et~al.}{2016}]{Garcia-Burillo2016}
{Garc{\'\i}a-Burillo} S.,  et~al., 2016, \mn@doi [\apjl]
  {10.3847/2041-8205/823/1/L12}, \href
  {https://ui.adsabs.harvard.edu/abs/2016ApJ...823L..12G} {823, L12}

\bibitem[\protect\citeauthoryear{{Garc{\'\i}a-Burillo}
  et~al.,}{{Garc{\'\i}a-Burillo} et~al.}{2019}]{Garcia-Burillo2019}
{Garc{\'\i}a-Burillo} S.,  et~al., 2019, \mn@doi [\aap]
  {10.1051/0004-6361/201936606}, \href
  {https://ui.adsabs.harvard.edu/abs/2019A&A...632A..61G} {632, A61}

\bibitem[\protect\citeauthoryear{Hafen et~al.,}{Hafen et~al.}{2019}]{Hafen2019}
Hafen Z.,  et~al., 2019, \mn@doi [\mnras] {10.1093/mnras/stz1773}, 488, 1248

\bibitem[\protect\citeauthoryear{{Hamann}, {Kanekar}, {Prochaska}, {Murphy},
  {Ellison}, {Malec}, {Milutinovic}  \& {Ubachs}}{{Hamann}
  et~al.}{2011}]{Hamann2011}
{Hamann} F.,  {Kanekar} N.,  {Prochaska} J.~X.,  {Murphy} M.~T.,  {Ellison} S.,
   {Malec} A.~L.,  {Milutinovic} N.,   {Ubachs} W.,  2011, \mn@doi [\mnras]
  {10.1111/j.1365-2966.2010.17575.x}, \href
  {https://ui.adsabs.harvard.edu/abs/2011MNRAS.410.1957H} {410, 1957}

\bibitem[\protect\citeauthoryear{{H{\"a}ring} \& {Rix}}{{H{\"a}ring} \&
  {Rix}}{2004}]{Haring2004}
{H{\"a}ring} N.,  {Rix} H.-W.,  2004, \mn@doi [\apjl] {10.1086/383567}, \href
  {https://ui.adsabs.harvard.edu/abs/2004ApJ...604L..89H} {604, L89}

\bibitem[\protect\citeauthoryear{{Harrison}, {Alexander}, {Mullaney}  \&
  {Swinbank}}{{Harrison} et~al.}{2014}]{Harrison2014}
{Harrison} C.~M.,  {Alexander} D.~M.,  {Mullaney} J.~R.,   {Swinbank} A.~M.,
  2014, \mn@doi [\mnras] {10.1093/mnras/stu515}, \href
  {https://ui.adsabs.harvard.edu/abs/2014MNRAS.441.3306H} {441, 3306}

\bibitem[\protect\citeauthoryear{{Hopkins}}{{Hopkins}}{2015}]{Hopkins2015}
{Hopkins} P.~F.,  2015, \mn@doi [\mnras] {10.1093/mnras/stv195}, \href
  {https://ui.adsabs.harvard.edu/abs/2015MNRAS.450...53H} {450, 53}

\bibitem[\protect\citeauthoryear{{Hopkins} \& {Quataert}}{{Hopkins} \&
  {Quataert}}{2011}]{Hopkins2011}
{Hopkins} P.~F.,  {Quataert} E.,  2011, \mn@doi [\mnras]
  {10.1111/j.1365-2966.2011.18542.x}, \href
  {https://ui.adsabs.harvard.edu/abs/2011MNRAS.415.1027H} {415, 1027}

\bibitem[\protect\citeauthoryear{{Hopkins}, {Torrey}, {Faucher-Gigu{\`e}re},
  {Quataert}  \& {Murray}}{{Hopkins} et~al.}{2016}]{Hopkins2016}
{Hopkins} P.~F.,  {Torrey} P.,  {Faucher-Gigu{\`e}re} C.-A.,  {Quataert} E.,
  {Murray} N.,  2016, \mn@doi [\mnras] {10.1093/mnras/stw289}, \href
  {https://ui.adsabs.harvard.edu/abs/2016MNRAS.458..816H} {458, 816}

\bibitem[\protect\citeauthoryear{{Hopkins} et~al.,}{{Hopkins}
  et~al.}{2018}]{Hopkins2018a}
{Hopkins} P.~F.,  et~al., 2018, \mn@doi [\mnras] {10.1093/mnras/sty674}, \href
  {https://ui.adsabs.harvard.edu/abs/2018MNRAS.477.1578H} {477, 1578}

\bibitem[\protect\citeauthoryear{{Imanishi}, {Nakanishi}  \&
  {Izumi}}{{Imanishi} et~al.}{2016}]{Imanishi2016}
{Imanishi} M.,  {Nakanishi} K.,   {Izumi} T.,  2016, \mn@doi [\apjl]
  {10.3847/2041-8205/822/1/L10}, \href
  {https://ui.adsabs.harvard.edu/abs/2016ApJ...822L..10I} {822, L10}

\bibitem[\protect\citeauthoryear{{Imanishi}, {Nakanishi}, {Izumi}  \&
  {Wada}}{{Imanishi} et~al.}{2018}]{Imanishi2018}
{Imanishi} M.,  {Nakanishi} K.,  {Izumi} T.,   {Wada} K.,  2018, \mn@doi
  [\apjl] {10.3847/2041-8213/aaa8df}, \href
  {https://ui.adsabs.harvard.edu/abs/2018ApJ...853L..25I} {853, L25}

\bibitem[\protect\citeauthoryear{{Imanishi} et~al.,}{{Imanishi}
  et~al.}{2020}]{Imanishi2020}
{Imanishi} M.,  et~al., 2020, \mn@doi [\apj] {10.3847/1538-4357/abaf50}, \href
  {https://ui.adsabs.harvard.edu/abs/2020ApJ...902...99I} {902, 99}

\bibitem[\protect\citeauthoryear{{Impellizzeri} et~al.,}{{Impellizzeri}
  et~al.}{2019}]{Impellizzeri2019}
{Impellizzeri} C.~M.~V.,  et~al., 2019, \mn@doi [\apjl]
  {10.3847/2041-8213/ab3c64}, \href
  {https://ui.adsabs.harvard.edu/abs/2019ApJ...884L..28I} {884, L28}

\bibitem[\protect\citeauthoryear{{Indriolo} \& {McCall}}{{Indriolo} \&
  {McCall}}{2012}]{Indriolo2012}
{Indriolo} N.,  {McCall} B.~J.,  2012, \mn@doi [\apj]
  {10.1088/0004-637X/745/1/91}, \href
  {https://ui.adsabs.harvard.edu/abs/2012ApJ...745...91I} {745, 91}

\bibitem[\protect\citeauthoryear{Iwamoto, Brachwitz, Nomoto, Kishimoto, Umeda,
  Hix  \& Thielemann}{Iwamoto et~al.}{1999}]{Iwamoto1999}
Iwamoto K.,  Brachwitz F.,  Nomoto K.,  Kishimoto N.,  Umeda H.,  Hix W.~R.,
  Thielemann F.-K.,  1999, \mn@doi [\apjs] {10.1086/313278}, 125, 439

\bibitem[\protect\citeauthoryear{Izzard, Tout, Karakas  \& Pols}{Izzard
  et~al.}{2004}]{Izzard2004}
Izzard R.~G.,  Tout C.~A.,  Karakas A.~I.,   Pols O.~R.,  2004, \mn@doi
  [\mnras] {10.1111/j.1365-2966.2004.07446.x}, 350, 407

\bibitem[\protect\citeauthoryear{{Kennicutt}}{{Kennicutt}}{1998}]{Kennicutt1998}
{Kennicutt} Robert~C. J.,  1998, \mn@doi [\apj] {10.1086/305588}, \href
  {https://ui.adsabs.harvard.edu/abs/1998ApJ...498..541K} {498, 541}

\bibitem[\protect\citeauthoryear{{Knapen}, {Comer{\'o}n}  \& {Seidel}}{{Knapen}
  et~al.}{2019}]{Knapen2019}
{Knapen} J.~H.,  {Comer{\'o}n} S.,   {Seidel} M.~K.,  2019, \mn@doi [\aap]
  {10.1051/0004-6361/201834669}, \href
  {https://ui.adsabs.harvard.edu/abs/2019A&A...621L...5K} {621, L5}

\bibitem[\protect\citeauthoryear{{Krajnovi{\'c}}, {Cappellari}, {de Zeeuw}  \&
  {Copin}}{{Krajnovi{\'c}} et~al.}{2006}]{Krajnovic2006}
{Krajnovi{\'c}} D.,  {Cappellari} M.,  {de Zeeuw} P.~T.,   {Copin} Y.,  2006,
  \mn@doi [\mnras] {10.1111/j.1365-2966.2005.09902.x}, \href
  {https://ui.adsabs.harvard.edu/abs/2006MNRAS.366..787K} {366, 787}

\bibitem[\protect\citeauthoryear{{Kroupa}}{{Kroupa}}{2002}]{Kroupa2002}
{Kroupa} P.,  2002, \mn@doi [Science] {10.1126/science.1067524}, \href
  {https://ui.adsabs.harvard.edu/abs/2002Sci...295...82K} {295, 82}

\bibitem[\protect\citeauthoryear{{Lagos}, {Padilla}, {Davis}, {Lacey}, {Baugh},
  {Gonzalez-Perez}, {Zwaan}  \& {Contreras}}{{Lagos} et~al.}{2015}]{Lagos2015}
{Lagos} C. d.~P.,  {Padilla} N.~D.,  {Davis} T.~A.,  {Lacey} C.~G.,  {Baugh}
  C.~M.,  {Gonzalez-Perez} V.,  {Zwaan} M.~A.,   {Contreras} S.,  2015, \mn@doi
  [\mnras] {10.1093/mnras/stu2763}, \href
  {https://ui.adsabs.harvard.edu/abs/2015MNRAS.448.1271L} {448, 1271}

\bibitem[\protect\citeauthoryear{Leitherer et~al.,}{Leitherer
  et~al.}{1999}]{Leitherer1999}
Leitherer C.,  et~al., 1999, \mn@doi [\apjs] {10.1086/313233}, 123, 3

\bibitem[\protect\citeauthoryear{{Li} \& {Goldsmith}}{{Li} \&
  {Goldsmith}}{2003}]{Li2003}
{Li} D.,  {Goldsmith} P.~F.,  2003, \mn@doi [\apj] {10.1086/346227}, \href
  {https://ui.adsabs.harvard.edu/abs/2003ApJ...585..823L} {585, 823}

\bibitem[\protect\citeauthoryear{{Lodato} \& {Bertin}}{{Lodato} \&
  {Bertin}}{2003}]{Lodato2003}
{Lodato} G.,  {Bertin} G.,  2003, \mn@doi [\aap] {10.1051/0004-6361:20021672},
  \href {https://ui.adsabs.harvard.edu/abs/2003A&A...398..517L} {398, 517}

\bibitem[\protect\citeauthoryear{Ma, Hopkins, {Faucher-Gigu{\`e}re}, Zolman,
  Muratov, Kere{\v s}  \& Quataert}{Ma et~al.}{2016}]{Ma2016}
Ma X.,  Hopkins P.~F.,  {Faucher-Gigu{\`e}re} C.-A.,  Zolman N.,  Muratov
  A.~L.,  Kere{\v s} D.,   Quataert E.,  2016, \mn@doi [\mnras]
  {10.1093/mnras/stv2659}, 456, 2140

\bibitem[\protect\citeauthoryear{Mannucci, Della~Valle  \& Panagia}{Mannucci
  et~al.}{2006}]{Mannucci2006}
Mannucci F.,  Della~Valle M.,   Panagia N.,  2006, \mn@doi [\mnras]
  {10.1111/j.1365-2966.2006.10501.x}, 370, 773

\bibitem[\protect\citeauthoryear{{Martin} \& {HIGGS Team}}{{Martin} \& {HIGGS
  Team}}{2011}]{Martin2011}
{Martin} C.~L.,  {HIGGS Team} 2011, in American Astronomical Society Meeting
  Abstracts \#217. p. 255.06

\bibitem[\protect\citeauthoryear{{McConnell} \& {Ma}}{{McConnell} \&
  {Ma}}{2013}]{McConnell2013}
{McConnell} N.~J.,  {Ma} C.-P.,  2013, \mn@doi [\apj]
  {10.1088/0004-637X/764/2/184}, \href
  {https://ui.adsabs.harvard.edu/abs/2013ApJ...764..184M} {764, 184}

\bibitem[\protect\citeauthoryear{{Meier} \& {Turner}}{{Meier} \&
  {Turner}}{2005}]{Meier2005}
{Meier} D.~S.,  {Turner} J.~L.,  2005, \mn@doi [\apj] {10.1086/426499}, \href
  {https://ui.adsabs.harvard.edu/abs/2005ApJ...618..259M} {618, 259}

\bibitem[\protect\citeauthoryear{{Moe}, {Arav}, {Bautista}  \& {Korista}}{{Moe}
  et~al.}{2009}]{Moe2009}
{Moe} M.,  {Arav} N.,  {Bautista} M.~A.,   {Korista} K.~T.,  2009, \mn@doi
  [\apj] {10.1088/0004-637X/706/1/525}, \href
  {https://ui.adsabs.harvard.edu/abs/2009ApJ...706..525M} {706, 525}

\bibitem[\protect\citeauthoryear{{M{\"u}ller-S{\'a}nchez}, {Prieto}, {Hicks},
  {Vives-Arias}, {Davies}, {Malkan}, {Tacconi}  \&
  {Genzel}}{{M{\"u}ller-S{\'a}nchez} et~al.}{2011}]{MuellerSanchez2011}
{M{\"u}ller-S{\'a}nchez} F.,  {Prieto} M.~A.,  {Hicks} E.~K.~S.,  {Vives-Arias}
  H.,  {Davies} R.~I.,  {Malkan} M.,  {Tacconi} L.~J.,   {Genzel} R.,  2011,
  \mn@doi [\apj] {10.1088/0004-637X/739/2/69}, \href
  {https://ui.adsabs.harvard.edu/abs/2011ApJ...739...69M} {739, 69}

\bibitem[\protect\citeauthoryear{Muratov, Kere{\v s}, {Faucher-Gigu{\`e}re},
  Hopkins, Quataert  \& Murray}{Muratov et~al.}{2015}]{Muratov2015}
Muratov A.~L.,  Kere{\v s} D.,  {Faucher-Gigu{\`e}re} C.-A.,  Hopkins P.~F.,
  Quataert E.,   Murray N.,  2015, \mn@doi [\mnras] {10.1093/mnras/stv2126},
  454, 2691

\bibitem[\protect\citeauthoryear{Muratov et~al.,}{Muratov
  et~al.}{2017}]{Muratov2017}
Muratov A.~L.,  et~al., 2017, \mn@doi [\mnras] {10.1093/mnras/stx667}, 468,
  4170

\bibitem[\protect\citeauthoryear{{Murray} \& {Chiang}}{{Murray} \&
  {Chiang}}{1995}]{Murray1995}
{Murray} N.,  {Chiang} J.,  1995, \mn@doi [\apjl] {10.1086/309775}, \href
  {https://ui.adsabs.harvard.edu/abs/1995ApJ...454L.105M} {454, L105}

\bibitem[\protect\citeauthoryear{{Narayanan}, {Krumholz}, {Ostriker}  \&
  {Hernquist}}{{Narayanan} et~al.}{2012}]{Narayanan2012}
{Narayanan} D.,  {Krumholz} M.~R.,  {Ostriker} E.~C.,   {Hernquist} L.,  2012,
  \mn@doi [\mnras] {10.1111/j.1365-2966.2012.20536.x}, \href
  {https://ui.adsabs.harvard.edu/abs/2012MNRAS.421.3127N} {421, 3127}

\bibitem[\protect\citeauthoryear{Nomoto, Tominaga, Umeda, Kobayashi  \&
  Maeda}{Nomoto et~al.}{2006}]{Nomoto2006}
Nomoto K.,  Tominaga N.,  Umeda H.,  Kobayashi C.,   Maeda K.,  2006, \mn@doi
  [Nuclear Physics A] {10.1016/j.nuclphysa.2006.05.008}, 777, 424

\bibitem[\protect\citeauthoryear{Orr et~al.,}{Orr et~al.}{2018}]{Orr2018}
Orr M.~E.,  et~al., 2018, \mn@doi [\mnras] {10.1093/mnras/sty1241}, 478, 3653

\bibitem[\protect\citeauthoryear{Pandya et~al.,}{Pandya
  et~al.}{2021}]{Pandya2021}
Pandya V.,  et~al., 2021, \mn@doi [\mnras] {10.1093/mnras/stab2714}, 508, 2979

\bibitem[\protect\citeauthoryear{{Raouf}, {Shabala}, {Croton}, {Khosroshahi}
  \& {Bernyk}}{{Raouf} et~al.}{2017}]{Raouf2017}
{Raouf} M.,  {Shabala} S.~S.,  {Croton} D.~J.,  {Khosroshahi} H.~G.,   {Bernyk}
  M.,  2017, \mn@doi [\mnras] {10.1093/mnras/stx1598}, \href
  {https://ui.adsabs.harvard.edu/abs/2017MNRAS.471..658R} {471, 658}

\bibitem[\protect\citeauthoryear{{Raouf}, {Khosroshahi}, {Mamon}, {Croton},
  {Hashemizadeh}  \& {Dariush}}{{Raouf} et~al.}{2018}]{Raouf2018}
{Raouf} M.,  {Khosroshahi} H.~G.,  {Mamon} G.~A.,  {Croton} D.~J.,
  {Hashemizadeh} A.,   {Dariush} A.~A.,  2018, \mn@doi [\apj]
  {10.3847/1538-4357/aace57}, \href
  {https://ui.adsabs.harvard.edu/abs/2018ApJ...863...40R} {863, 40}

\bibitem[\protect\citeauthoryear{{Raouf}, {Silk}, {Shabala}, {Mamon}, {Croton},
  {Khosroshahi}  \& {Beckmann}}{{Raouf} et~al.}{2019}]{Raouf2019}
{Raouf} M.,  {Silk} J.,  {Shabala} S.~S.,  {Mamon} G.~A.,  {Croton} D.~J.,
  {Khosroshahi} H.~G.,   {Beckmann} R.~S.,  2019, \mn@doi [\mnras]
  {10.1093/mnras/stz907}, \href
  {https://ui.adsabs.harvard.edu/abs/2019MNRAS.486.1509R} {486, 1509}

\bibitem[\protect\citeauthoryear{{Raouf} et~al.,}{{Raouf}
  et~al.}{2021}]{Raouf2021}
{Raouf} M.,  et~al., 2021, \mn@doi [\apj] {10.3847/1538-4357/abd47d}, \href
  {https://ui.adsabs.harvard.edu/abs/2021ApJ...908..123R} {908, 123}

\bibitem[\protect\citeauthoryear{{Richings}, {Schaye}  \&
  {Oppenheimer}}{{Richings} et~al.}{2014a}]{Richings2014a}
{Richings} A.~J.,  {Schaye} J.,   {Oppenheimer} B.~D.,  2014a, \mn@doi [\mnras]
  {10.1093/mnras/stu525}, \href
  {https://ui.adsabs.harvard.edu/abs/2014MNRAS.440.3349R} {440, 3349}

\bibitem[\protect\citeauthoryear{{Richings}, {Schaye}  \&
  {Oppenheimer}}{{Richings} et~al.}{2014b}]{Richings2014b}
{Richings} A.~J.,  {Schaye} J.,   {Oppenheimer} B.~D.,  2014b, \mn@doi [\mnras]
  {10.1093/mnras/stu1046}, \href
  {https://ui.adsabs.harvard.edu/abs/2014MNRAS.442.2780R} {442, 2780}

\bibitem[\protect\citeauthoryear{{Richings}, {Faucher-Gigu{\`e}re}  \&
  {Stern}}{{Richings} et~al.}{2021}]{Richings2021}
{Richings} A.~J.,  {Faucher-Gigu{\`e}re} C.-A.,   {Stern} J.,  2021, \mn@doi
  [\mnras] {10.1093/mnras/stab556}, \href
  {https://ui.adsabs.harvard.edu/abs/2021MNRAS.503.1568R} {503, 1568}

\bibitem[\protect\citeauthoryear{{Roy}, {Wilson}, {Ulvestad}  \&
  {Colbert}}{{Roy} et~al.}{2000}]{Roy2000}
{Roy} A.~L.,  {Wilson} A.~S.,  {Ulvestad} J.~S.,   {Colbert} J.~M.,  2000, in
  {Conway} J.~E.,  {Polatidis} A.~G.,  {Booth} R.~S.,   {Pihlstr{\"o}m} Y.~M.,
  eds, EVN Symposium 2000, Proceedings of the 5th european VLBI Network
  Symposium. p.~7 (\mn@eprint {arXiv} {astro-ph/0009408}),
  \mn@doi{10.48550/arXiv.astro-ph/0009408}

\bibitem[\protect\citeauthoryear{{Sch{\"o}ier}, {van der Tak}, {van Dishoeck}
  \& {Black}}{{Sch{\"o}ier} et~al.}{2005}]{Schoier2005}
{Sch{\"o}ier} F.~L.,  {van der Tak} F.~F.~S.,  {van Dishoeck} E.~F.,   {Black}
  J.~H.,  2005, \mn@doi [\aap] {10.1051/0004-6361:20041729}, \href
  {https://ui.adsabs.harvard.edu/abs/2005A&A...432..369S} {432, 369}

\bibitem[\protect\citeauthoryear{{Scourfield} et~al.,}{{Scourfield}
  et~al.}{2020}]{Scourfield2020}
{Scourfield} M.,  et~al., 2020, \mn@doi [\mnras] {10.1093/mnras/staa1891},
  \href {https://ui.adsabs.harvard.edu/abs/2020MNRAS.496.5308S} {496, 5308}

\bibitem[\protect\citeauthoryear{{Shin}, {Woo}, {Kim}  \& {Wang}}{{Shin}
  et~al.}{2021}]{Shin2021}
{Shin} J.,  {Woo} J.-H.,  {Kim} M.,   {Wang} J.,  2021, \mn@doi [\apj]
  {10.3847/1538-4357/abd779}, \href
  {https://ui.adsabs.harvard.edu/abs/2021ApJ...908...81S} {908, 81}

\bibitem[\protect\citeauthoryear{{Silk} \& {Rees}}{{Silk} \&
  {Rees}}{1998}]{Silk1998}
{Silk} J.,  {Rees} M.~J.,  1998, \aap, \href
  {https://ui.adsabs.harvard.edu/abs/1998A&A...331L...1S} {331, L1}

\bibitem[\protect\citeauthoryear{{Silpa}, {Kharb}, {O'Dea}, {Baum},
  {Sebastian}, {Mukherjee}  \& {Harrison}}{{Silpa} et~al.}{2021}]{Silpa2021}
{Silpa} S.,  {Kharb} P.,  {O'Dea} C.~P.,  {Baum} S.~A.,  {Sebastian} B.,
  {Mukherjee} D.,   {Harrison} C.~M.,  2021, \mn@doi [\mnras]
  {10.1093/mnras/stab2110}, \href
  {https://ui.adsabs.harvard.edu/abs/2021MNRAS.507.2550S} {507, 2550}

\bibitem[\protect\citeauthoryear{{Silpa}, {Kharb}, {Harrison}, {Girdhar},
  {Mukherjee}, {Mainieri}  \& {Jarvis}}{{Silpa} et~al.}{2022}]{Silpa2022}
{Silpa} S.,  {Kharb} P.,  {Harrison} C.~M.,  {Girdhar} A.,  {Mukherjee} D.,
  {Mainieri} V.,   {Jarvis} M.~E.,  2022, \mn@doi [\mnras]
  {10.1093/mnras/stac1044}, \href
  {https://ui.adsabs.harvard.edu/abs/2022MNRAS.513.4208S} {513, 4208}

\bibitem[\protect\citeauthoryear{{Springel}, {Di Matteo}  \&
  {Hernquist}}{{Springel} et~al.}{2005}]{Springel2005a}
{Springel} V.,  {Di Matteo} T.,   {Hernquist} L.,  2005, \mn@doi [\mnras]
  {10.1111/j.1365-2966.2005.09238.x}, \href
  {https://ui.adsabs.harvard.edu/abs/2005MNRAS.361..776S} {361, 776}

\bibitem[\protect\citeauthoryear{{Torrey} et~al.,}{{Torrey}
  et~al.}{2020}]{Torrey2020}
{Torrey} P.,  et~al., 2020, \mn@doi [\mnras] {10.1093/mnras/staa2222}, \href
  {https://ui.adsabs.harvard.edu/abs/2020MNRAS.497.5292T} {497, 5292}

\bibitem[\protect\citeauthoryear{{Viti} et~al.,}{{Viti}
  et~al.}{2014}]{Viti2014}
{Viti} S.,  et~al., 2014, \mn@doi [\aap] {10.1051/0004-6361/201424116}, \href
  {https://ui.adsabs.harvard.edu/abs/2014A&A...570A..28V} {570, A28}

\bibitem[\protect\citeauthoryear{{Vollmer} \& {Davies}}{{Vollmer} \&
  {Davies}}{2013}]{Vollmer2013}
{Vollmer} B.,  {Davies} R.~I.,  2013, \mn@doi [\aap]
  {10.1051/0004-6361/201321409}, \href
  {https://ui.adsabs.harvard.edu/abs/2013A&A...556A..31V} {556, A31}

\bibitem[\protect\citeauthoryear{{Vollmer} et~al.,}{{Vollmer}
  et~al.}{2022}]{Vollmer2022}
{Vollmer} B.,  et~al., 2022, arXiv e-prints, \href
  {https://ui.adsabs.harvard.edu/abs/2022arXiv220614513V} {p. arXiv:2206.14513}

\bibitem[\protect\citeauthoryear{{Watanabe}, {Sakai}, {Sorai}  \&
  {Yamamoto}}{{Watanabe} et~al.}{2014}]{Watanabe2014}
{Watanabe} Y.,  {Sakai} N.,  {Sorai} K.,   {Yamamoto} S.,  2014, \mn@doi [\apj]
  {10.1088/0004-637X/788/1/4}, \href
  {https://ui.adsabs.harvard.edu/abs/2014ApJ...788....4W} {788, 4}

\bibitem[\protect\citeauthoryear{{Winkel} et~al.,}{{Winkel}
  et~al.}{2022}]{Winkel2022}
{Winkel} N.,  et~al., 2022, \mn@doi [\aap] {10.1051/0004-6361/202243697}, \href
  {https://ui.adsabs.harvard.edu/abs/2022A&A...663A.104W} {663, A104}

\bibitem[\protect\citeauthoryear{{Zakamska} et~al.,}{{Zakamska}
  et~al.}{2016}]{Zakamska2016}
{Zakamska} N.~L.,  et~al., 2016, \mn@doi [\mnras] {10.1093/mnras/stw718}, \href
  {https://ui.adsabs.harvard.edu/abs/2016MNRAS.459.3144Z} {459, 3144}

\bibitem[\protect\citeauthoryear{{Zubovas} \& {King}}{{Zubovas} \&
  {King}}{2014}]{Zubovas2014}
{Zubovas} K.,  {King} A.~R.,  2014, \mn@doi [\mnras] {10.1093/mnras/stt2472},
  \href {https://ui.adsabs.harvard.edu/abs/2014MNRAS.439..400Z} {439, 400}

\bibitem[\protect\citeauthoryear{{von Steiger} \& {Zurbuchen}}{{von Steiger} \&
  {Zurbuchen}}{2016}]{vonSteiger2016}
{von Steiger} R.,  {Zurbuchen} T.~H.,  2016, \mn@doi [\apj]
  {10.3847/0004-637X/816/1/13}, \href
  {https://ui.adsabs.harvard.edu/abs/2016ApJ...816...13V} {816, 13}

\makeatother
\end{thebibliography}

\label{lastpage}
\end{document}